\documentclass[aps,prl,reprint,superscriptaddress]{revtex4-1}
\usepackage{amsmath}
\usepackage{graphicx}
\usepackage{amsfonts}
\usepackage{color}
\usepackage{upgreek}
\usepackage{bm}
\usepackage{lipsum}
\usepackage{comment}

\usepackage[table]{xcolor}
\usepackage{textcomp}
\usepackage{amsmath}
\usepackage{amssymb}
\usepackage{graphicx} 
\usepackage{pgf}
\usepackage{epstopdf}
\usepackage{braket}
\usepackage{mathtools}
\usepackage{times}

\usepackage[retainorgcmds]{IEEEtrantools}

\newcommand{\ee}{\mathrm{e}}
\newcommand{\ii}{\mathrm{i}}

\newcommand{\Tr}{\mathrm{Tr}}

\newcommand*{\defeq}{\mathrel{\vcenter{\baselineskip0.5ex\lineskiplimit0pt\hbox{\scriptsize.}\hbox{\scriptsize.}}}=}
\newcommand*{\defeqr}{=\mathrel{\vcenter{\baselineskip0.5ex\lineskiplimit0pt\hbox{\scriptsize.}\hbox{\scriptsize.}}}}
\newcommand{\vvar}{\textrm{var}}
\newcommand{\ccov}{\textrm{cov}}

\usepackage[colorlinks,linkcolor=blue,urlcolor=blue,citecolor=blue]{hyperref}

\begin{document}

\title{Particle current statistics in driven mesoscale conductors}

\author{Marlon Brenes}
\email{marlon.brenes@utoronto.ca}
\affiliation{Department of Physics and Centre for Quantum Information and Quantum Control, University of Toronto, 60 Saint George St., Toronto, Ontario, M5S 1A7, Canada}

\author{Giacomo Guarnieri}
\affiliation{Dahlem Centre for Complex Quantum Systems, Freie Universität Berlin, 14195 Berlin, Germany}

\author{Archak Purkayastha}
\affiliation{School of Physics, Trinity College Dublin, College Green, Dublin 2, Ireland}
\affiliation{Centre for Complex Quantum Systems, Department of Physics and Astronomy, Aarhus University, Ny Munkegade 120, DK-8000 Aarhus C, Denmark}

\author{Jens Eisert}
\affiliation{Dahlem Centre for Complex Quantum Systems, Freie Universität Berlin, 14195 Berlin, Germany}

\author{Dvira Segal}
\affiliation{Department of Chemistry, University of Toronto, 80 Saint George St., Toronto, Ontario, M5S 3H6, Canada}

\affiliation{Department of Physics and Centre for Quantum Information and Quantum Control, University of Toronto, 60 Saint George St., Toronto, Ontario, M5S 1A7, Canada}

\author{Gabriel Landi}
\affiliation{Instituto de F\'isica da Universidade de S\~ao Paulo, 05314-970 S\~ao Paulo, Brazil}

\affiliation{Department of Physics and Astronomy, University of Rochester, Rochester, New York 14627, USA}
  
\date{\today}

\begin{abstract}
We propose a highly-scalable method to compute the statistics of charge transfer in driven conductors. The framework can be applied  in situations of non-zero temperature, strong coupling to terminals and in the presence of non-periodic light-matter interactions, away from equilibrium. The approach combines the so-called mesoscopic leads formalism with full counting statistics. It results in a generalised quantum master equation that dictates the dynamics of current fluctuations and higher order moments of the probability distribution function of charge exchange. For generic time-dependent quadratic Hamiltonians, we provide closed-form expressions for computing noise in the non-perturbative regime of the parameters of the system, reservoir or system-reservoir interactions. Having access to the full dynamics of the current and its noise, the method allows us to compute the variance of charge transfer over time in non-equilibrium configurations. The dynamics reveal that in driven systems, the average noise should be defined operationally with care over which period of time is covered. 
\end{abstract}

\maketitle
Current fluctuations are inherent to out-of-equilibrium mesoscopic devices operating in the quantum regime~\cite{EspositoRev2009,Gustavsson2006,Flindt2009,Fricke2010,Ubbelohde2012}. Their categorisation and quantification is relevant to the understanding of fundamental thermodynamics as well as the operation of quantum thermal machines~\cite{LandiRev2021,Goold2016,Binder2018,christian_review}. 
Recent experimental advances include non-periodic modulation in light-induced currents~\cite{Boolakee2022,Khan2021} and the control of the system-reservoir interactions in superconducting circuits~\cite{Ronzani2018}
as well as single-molecule junctions~\cite{Herre,Bin}. These advances call for methodologies that allow one to cope with the effects of these physical properties at finite-temperature to understand fluctuations in their regimes of operation.

Most of the existing methods for computing current fluctuations, however, are only applicable in restricted regimes of operation. When solely coherent quantum effects are important and there is no time-dependence in the Hamiltonian, the 
\emph{Levitov-Lesovik approach} \cite{Levitov1993,Levitov1996}, which extends the \emph{Landauer scattering theory}~\cite{Blanter2001}, provides non-perturbative exact results. \emph{Green's function techniques} can be formulated to treat strong system-reservoir coupling~\cite{Misha}. However, to include in this approach either a 
time-periodic drive or incoherent effects arising from many-body interactions typically requires treatments via non-equilibrium Green's functions~\cite{Hanggi2003,Stefanucci_2009,Wang_2013,Talarico:2020}. These methods are perturbative either in the Hamiltonian parameters or the drive parameters, and naturally cannot be applied to cases lacking a perturbative parameter, such as when applying a strong non-periodic drive on the nanoscale conductor.

If the system-reservoir coupling energy is weak, \emph{quantum master equations} (QMEs) offer an alternative, flexible route to evaluate both average currents and their fluctuations~\cite{EspositoRev2009,LandiRev2021_2,Benito:2016,Friedman_2018,Liu:2021}.

While for small systems, QME methods can handle many-body interactions in a non-perturbative manner, these methods are fundamentally limited in their ability to accurately and consistently describe the system's quantum state~\cite{Tupkary:2022}. 
Furthermore, it has been recently argued that an appropriate thermodynamic description at the {\em fluctuating} level may only be obtained after applying the secular approximation on the \emph{Redfield QME}~\cite{Soret:2022}.

We introduce a novel method that allows for the non-perturbative characterisation of current fluctuations in out-of-equilibrium configurations for arbitrarily-driven systems, overcoming the aforementioned limitations.
Our scalable method combines a \emph{full-counting statistics} (FCS) treatment~\cite{EspositoRev2009} with the so-called \emph{mesoscopic leads description}~\cite{Imamoglu1994,Garraway1997a,Garraway1997b,Subotnik:2009}
and brings together advantages of both approaches. Mesoscopic leads build the reservoirs by a finite collection of fermionic modes, each of which is subject to damping, intended to bring the discrete modes of the bath to their equilibrium state with respect to a fixed temperature and chemical potential. In its basic form, the mesoscopic leads approach has been shown to build the correct thermodynamic state~\cite{Gruss2016,Guimares2016,Elenewski2017,Uzdin2018,Reichental2018,Chen2019,Brenes:2020,Lacerda:2022} and it has been adopted to study non-interacting~\cite{Dzhioev2011,Ajisaka2012,Ajisaka2013,Zelovich2014}, periodically-driven~\cite{Chen:2014,Oz:2020,Lacerda:2022} and impurity~\cite{Schwarz2016,Schwarz2018,Lotem:2020} models, as well as thermal machines~\cite{Brenes:2020} in the strongly-interacting, finite-temperature and strong system-reservoir coupling regimes, away from equilibrium. 
The method presented in this work bridges and combines in a non-trivial way two established but separate frameworks, namely the mesoscopic leads approach and the FCS, yielding the charge current and its fluctuations for arbitrary system-reservoir coupling strength, temperature, bias-voltage and time-dependent driving fields. Further, in the case of Gaussian time-dependent quantum systems, our framework leads to elegant expressions for the instantaneous dynamics of the currents and its noise. Studying as an example a periodically-driven system, we compute the instantaneous charge current noise. We reveal the subtle nature of fluctuations under driving far away from equilibrium with the noise showing crucial dependency on the time interval under investigation. 

\begin{figure}[t]
\fontsize{13}{10}\selectfont 
\centering
\includegraphics[width=1\columnwidth]{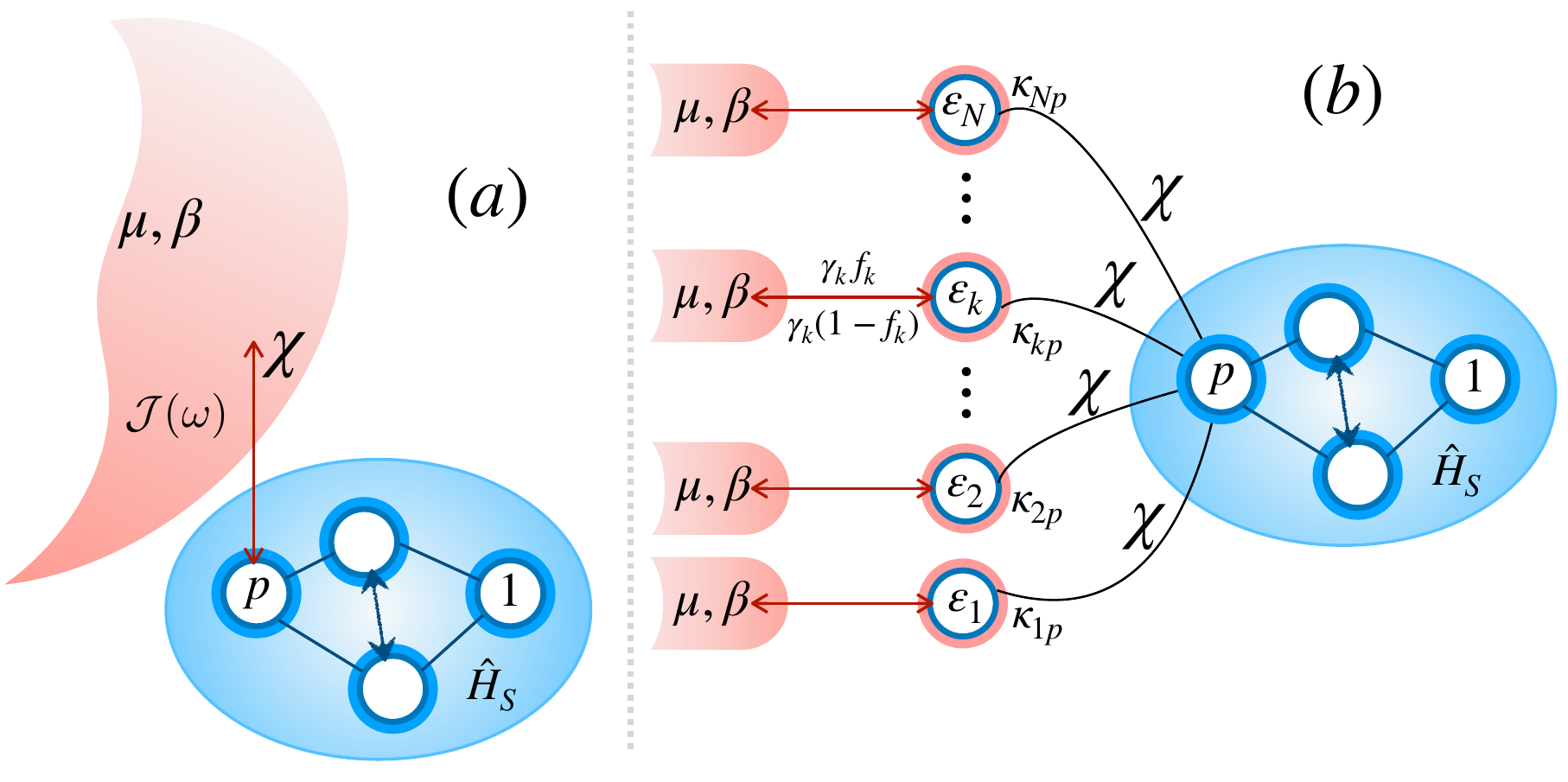}
\caption{Mesoscopic lead description of an open quantum system.
(a) A depiction of an infinite bath at temperature $T$ and chemical potential $\mu$ with spectral density function $\mathcal{J}(\omega)$ coupled locally to the $p$-th fermionic site of a system. 
(b) The bath is discretised by a finite collection of $N$ fermionic modes with self-energies $\varepsilon_{k}$, which are coupled locally to the $p$-th site of the system with strength $\kappa_{k,p}$. Each of the modes is subject to dissipation intended to drive the mode to thermal and chemical equilibrium state. In (a), FCS is performed with a counting field $\chi$ {\it embedded in the reservoir}. In contrast, in (b) the counting fields turn up in the {\it internal} system-modes couplings.}
\label{fig:1}
\end{figure}

\emph {Mesoscopic reservoirs}. We consider a fermionic system ${\tt S}$ described by a set of $L$ annihilation operators $\{\hat{c}_j\}$ and a Hamiltonian $\hat{H}_{\tt S}(t)$, possibly interacting and driven. 
The system is coupled to $Q$ fermionic reservoirs, each modelled by a set of operators 
$\{\hat{b}_{n,\alpha}\}$, Hamiltonians  $\hat{H}_{\tt B , \alpha} = \sum_{n=1}^{\infty} \omega_{n,\alpha} \hat{b}^{\dagger}_{n,\alpha} \hat{b}_{n,\alpha}$ (we set $\hbar = 1, k_{\rm B} = 1$) and prepared in grand-canonical states at temperatures $T_\alpha$ and chemical potentials $\mu_\alpha$.
Each reservoir $\alpha$ is assumed to couple to a specific system operator $\hat{c}_{p_\alpha}$ via
$\hat{H}_{\tt SB_\alpha} = \sum_{n=1}^{\infty} \lambda_{n, \alpha} 
( \hat{c}_{p_\alpha}^\dagger \hat{b}_{n,\alpha} + 
\hat{b}_{n,\alpha}^\dagger \hat{c}_{p_\alpha})$, 
which is not necessarily weak.
The corresponding bath spectral densities are $\mathcal{J}_\alpha(\omega) = 2\pi \sum_{n=1}^{\infty} |\lambda_{n,\alpha}|^2 \delta(\omega - \omega_{n, \alpha})$.
The combination of time-dependent drives and/or interactions in $t \mapsto \hat{H}_{\tt S}(t)$, together with strong couplings between the system and the fermionic baths, makes the above problem notoriously difficult to handle. 

The mesoscopic leads approach has been successful in this regard~\cite{Imamoglu1994,Garraway1997a,Garraway1997b,Zwolak1:2021,Zwolak2:2021,Brenes:2020,Lacerda:2022,Chen:2014}. Here, each reservoir $\alpha$ is mapped into a \emph{finite} set of $N_\alpha$ \emph{lead modes} $\{\hat{a}_{k,\alpha}\}$, $k = 1,\ldots, N_\alpha$, each of which is coupled to a residual reservoir, as depicted in Fig.~\ref{fig:1}. 
The method is designed so as it converges to the true dynamics when $N_\alpha\to\infty$.

The Hamiltonian of the leads reads $\hat{H}_{\tt{L}} = \sum_\alpha \sum_{k=1}^{N_{\alpha}} \varepsilon_{k , \alpha} \hat{a}^\dagger_{k , \alpha} \hat{a}_{k , \alpha}$, with each lead mode $\hat{a}_{k,\alpha}$  assigned an energy $\varepsilon_{k,\alpha}$, designed to homogeneously sample the spectral bandwidth of $\hat{H}_{{\tt B}_\alpha}$.
Moreover, ${\tt S}$ only interacts with the lead modes, and not their residual reservoirs. It follows that $\hat{H}_{\tt SB_{\alpha}} \mapsto \hat{H}_{\tt SL_{\alpha}} = \sum_{k=1}^{N_\alpha} \kappa_{k,\alpha} \big( \hat{c}_{p_\alpha}^\dagger \hat{a}_{k,\alpha} + \hat{a}_{k,\alpha}^\dagger \hat{c}_{p_\alpha}\big)$, with new coupling strengths $\kappa_{k,\alpha} = \sqrt{\mathcal{J}_\alpha(\varepsilon_{k,\alpha}) \gamma_{k,\alpha}/(2\pi)}$, where $\gamma_{k,\alpha} = \varepsilon_{k+1,\alpha}  -\varepsilon_{k,\alpha}$ will be small whenever $N_\alpha$ is large.
Crucially, via this mapping the residual environment of each lead mode has a flat spectral density, governed by $\gamma_{k,\alpha}$~\cite{Garraway1997a,Garraway1997b}. 
Thus, even if the original ${\tt SB}$ coupling is not weak, the coupling of the lead modes to their residual baths becomes small, provided $N_\alpha$ is sufficiently large~\footnote{More quantitatively, the condition is that $\gamma_{k , \alpha}$ needs to remain the smallest energy scale in the problem~\cite{Zwolak1:2021,Zwolak2:2021}.}. This condition allows one 
to trace out the residual environments and obtain a master equation for the joint system-state $\hat{\rho}_{\tt SL}$. 

\emph{Full counting statistics}.
The mesoscopic leads approach only gives access to average currents through continuity equations.
Our goal is to take this method a step further and construct  the full probability distribution of charge fluctuations.
Letting $I_\nu(t)$ denote the stochastic charge current to reservoir $\nu$ and $N_\nu(t,t_0)=\int_{t_0}^t {\rm d}t'~I_\nu(t')$ the corresponding integrated (net) charge in the interval $[t_0,t]$, our interest will be on the probability $P(n,t,t_0) = P\big(N_\nu(t,t_0) = n\big)$.
We have that~\cite{EspositoRev2009}
\begin{equation}\label{FCS_full_dist}
    P(n,t,t_0) = \int_{-\pi}^\pi \frac{{\rm d}\chi}{2\pi} e^{-\ii n\chi} G(\chi, t, t_0).
\end{equation}
As one of our main results, we show in the Supplemental Material~\cite{SM} that $G(\chi, t, t_0) \defeq \Tr[\hat{\rho}_{\tt SL}(\chi,t,t_0)]$ and $\hat{\rho}_{\tt SL}(\chi,t,t_0)$ satisfies the generalised master equation 
$\frac{\rm{d}}{{\rm{d}}t} \hat{\rho}_{\tt SL}(\chi,t,t_0) = \mathcal{L}_\chi(t) \hat{\rho}_{\tt SL}(\chi,t,t_0)$, 
with the tilted Liouvillian 
\begin{equation}
\label{tilted_L}
    \mathcal{L}_\chi(t) \hat{\rho} = -\ii [\hat{H}_{\tt S}(t) + \hat{H}_{\tt L} + \hat{H}_{\tt SL}^\chi, \hat{\rho}]_\chi + \sum_\alpha \mathcal{D}_\alpha \hat{\rho}.
\end{equation}
Here, $\chi$ is the counting field,
$[\hat{A}_\chi,\hat{B}]_\chi \defeq \hat{A}_\chi \hat{B} - \hat{B} \hat{A}_{-\chi}$,  
\begin{equation}
\label{tilted_H_SL}
    \hat{H}_{\tt SL}^\chi = \sum_{\alpha=1}^{Q} \sum_{k=1}^{N_\alpha} \kappa_{k,\alpha} \big( \hat{c}_{p_\alpha}^\dagger \hat{a}_{k,\alpha}~ e^{-i \chi \delta_{\alpha ,\nu}/2} + \hat{a}_{k,\alpha}^\dagger \hat{c}_{p_\alpha}~e^{i \chi \delta_{\alpha , \nu}/2}\big)
\end{equation}
and
\begin{eqnarray}
\label{eq:dissipator}
\mathcal{D}_{\alpha}\hat{\rho} &=& \sum_{k=1}^{N_\alpha} \gamma_{k , \alpha}(1 - f_{k , \alpha}) \left[\hat{a}_{k , \alpha} \hat{\rho} \hat{a}^{\dagger}_{k , \alpha} - \tfrac{1}{2}\{ \hat{a}^{\dagger}_{k , \alpha} \hat{a}_{k , \alpha}, \hat{\rho} \} \right] \nonumber \\
&+& \sum_{k=1}^{N_\alpha} \gamma_{k , \alpha} f_{k , \alpha} \left[\hat{a}^{\dagger}_{k , \alpha} \hat{\rho} \hat{a}_{k , \alpha} - \tfrac{1}{2}\{ \hat{a}_{k , \alpha} \hat{a}^{\dagger}_{k , \alpha}, \hat{\rho} \} \right].
\end{eqnarray}
The Lindblad dissipators $\mathcal{D}_{\alpha}$ are generators of quantum dynamical semi-groups: It is important to note that in this picture, they are made time-independent.
They act only locally on the individual lead modes 
$\hat{a}_{k,\alpha}$, with strength $\gamma_{k,\alpha}$ and Fermi-Dirac occupation $f_{k , \alpha} \defeq  (\ee^{(\varepsilon_{k , \alpha}-\mu_{\alpha})/T_\alpha} + 1)^{-1}$.
Setting $\chi=0$, one recovers the traditional mesoscopic leads master equation~\cite{Brenes:2020}.
With $\mathcal{L}_\chi$, however, we now have access to the full $P(n,t,t_0)$. Note that we have included an explicit dependence on the initial condition at $t_0$. As we shall see, it is important to keep track of this argument to evaluate charge statistics in systems with an explicit time-dependent Hamiltonian.
The counting field $\chi$ specifies which physical process we are monitoring. 
Charge transport is usually associated with quantum jumps in the master equation, with $\chi$ placed in the terms $\hat{a}_{k,\alpha} \hat{\rho} \hat{a}_{k,\alpha}^\dagger$ and $\hat{a}_{k,\alpha}^\dagger \hat{\rho} \hat{a}_{k,\alpha}$ of Eq.~\eqref{eq:dissipator}.
Instead, a crucial aspect of our result~\eqref{tilted_L} is that $\chi$ is placed in the unitary system-leads interactions, $\hat{c}_p^\dagger \hat{a}_{k,\alpha}$ and  $\hat{a}_{k,\alpha}^\dagger \hat{c}_p$. 
This is a consequence of the mapping, which implies that the exchange of particles between ${\tt S}$ and ${\tt B}$ is mapped to an exchange between ${\tt S}$ and the lead modes $\hat{a}_{k,\alpha}$. 
In non-driven systems at steady-state, such a distinction is immaterial. However, for driven systems, and during transients, it is crucial.

We note that the proposed scheme is based on the two-point measurement protocol FCS~\cite{EspositoRev2009} which can be justified with the assumption that initial total density matrix is a product state of system and environment states (see Ref.~\cite{SM} for further details).

\emph {Noise}. The average current, 
$J_\nu(t) \defeq \langle I_\nu(t)\rangle = \frac{{\rm d}}{{\rm d}t} \langle N_\nu(t,t_0) \rangle$ is given by~\cite{Brenes:2020,Lacerda:2022}
\begin{align}\label{Jave}
    J_{\nu}(t) = \ii \sum_{k=1}^{N_{\nu}} \kappa_{k,\nu} \Tr\big\{\big( \hat{c}^\dagger_{p_\nu} \hat{a}_{k,\nu} - \hat{a}_{k,\nu}^\dagger \hat{c}_{p_\nu} \big) \hat{\rho}_{\tt SL}(\chi=0,t,t_0)\big\},
\end{align}
and therefore does not require the tilted dynamics.
For all higher order moments, however, $\mathcal{L}_{\chi}$ is required.
Here, we focus on the charge variance  $\vvar[N(t,t_0)] \defeq \langle N^2_\nu(t,t_0) \rangle - \langle N_\nu(t,t_0) \rangle^2$ or, more conveniently, the \emph{noise}
\begin{equation}\label{noise}
    D_\nu(t,t_0) \defeq \frac{\rm d}{{\rm d}t} \vvar[N_\nu(t,t_0)]
    =2 \int_{t_0}^t {\rm d}t'~ \big\langle 
    \delta I_\nu(t) \delta I_\nu(t')
    \big\rangle,
\end{equation}
where $\delta I_\nu(t) = I_\nu(t) - J_\nu(t)$ and the last equality follows from $N_\nu(t,t_0) = \int_{t_0}^t {\rm{d}}t'~I_\nu(t')$.

A major advantage of our approach is the ability to describe arbitrary drives and transient dynamics. 
In such cases, it is crucial to note that while $J_\nu(t)$ is an instantaneous quantity,  $D_\nu(t,t_0)$ depends on the time 
interval $[t_0,t]$ in question.
At the stochastic level the charge is additive
as $N_\nu(t_2,t_0) = N_\nu(t_2,t_1) + N_\nu(t_1,t_0), \forall t_2 > t_1 > t_0$. In contrast, the variance {\it is not additive} since $\vvar(A+B) = \vvar(A) + \vvar(B) + 2\ccov(A,B)$. Eq.~\eqref{noise} thus  yields
\begin{equation}\label{Ds_covariance}
    D_\nu(t_2,t_0) = D_\nu(t_2,t_1) +2 \frac{{\rm d}}{{\rm d}t_2}\ccov\big[N_{\nu}(t_2,t_1),N_{\nu}(t_1,t_0)\big],
\end{equation}
which shows a dependence on the correlation between the transferred charge at different intervals. 
For systems with autonomous steady-states, it suffices to work with $\lim_{t\to\infty} D_\nu(t,t_0)$, and no such subtlety arises. However, this is not the case in driven systems. For example, in the case of periodic drives (with characteristic driving period $\tau$), $D_\nu(t_0+\tau,t_0)$ reflects fluctuations over a single period while $\lim_{t\to\infty} D_\nu(t,t_0)$ portrays the fluctuations over many periods. To our knowledge, there is currently no method capable to account for this distinction, and demonstrate its ramifications. 

\emph {Gaussian states and dynamics}. 
Our description thus far has made no assumption about the structure of $\hat{H}_{\tt S}(t)$. 
Arbitrary interacting systems are accessible and can be simulated using, e.g., \emph{tensor networks}, as put forth in Ref.~\cite{Brenes:2020}.
However, if $\hat{H}_{\tt S}(t)$ is quadratic in fermionic operators, the tilted Liouvillian~\eqref{tilted_L} is Gaussian-preserving.
Let $\{\hat{b}_i\} = \{\hat{c}_j, \hat{a}_{k,\alpha}\}$ denote a combined set of fermionic operators of the system plus the $Q$ leads. 
A quadratic $\hat{H}_{\tt S}(t)$  implies that we may write
$\hat{H}(t) = \hat{H}_{\tt S}(t) + \sum_{\alpha = 1}^{Q} ( \hat{H}_{\tt{L}_{\alpha}} + \hat{H}_{\tt{S}\tt{L}_{\alpha}}) \defeq \sum_{i,j} h_{i,j}(t) \hat{b}_i^{\dagger} \hat{b}_j$, for a matrix $\mathbf{H}(t)$ with matrix elements $h_{i,j}(t)$ of dimension $L+\sum_{\alpha=1}^Q N_\alpha$.
In the untilted case ($\chi=0)$, it is well-known that the particle-number preserving \emph{covariance matrix} $\mathbf{C}\geq 0$ with entries $[\mathbf{C}(t)]_{i,j} \defeq \Tr[\hat{b}_j^{\dagger} \hat{b}_i \hat{\rho}(t)]$  evolves according to the Lyapunov equation~\cite{Purkayastha2022,Lacerda:2022,Prosen:2008}
\begin{align}
\label{eq:lyapunov}
    \frac{{\rm{d}}\mathbf{C}(t)}{{\rm{d}}t} 
     &= - \left[\mathbf{W}(t)\mathbf{C}(t) + \mathbf{C}(t) \mathbf{W}^\dagger(t) \right] + \mathbf{F},
\end{align}
where $[\mathbf{W}(t)]_{i,j} = \ii h_{i,j}(t) + \gamma_{i,j}/2$ and 
$\bm{\gamma}$
is a diagonal matrix with entries $\gamma_{k,\alpha}$ [Eq.~\eqref{eq:dissipator}] in the sector of the leads.
Similarly, $\mathbf{F}$ is a diagonal matrix with entries $\gamma_{k,\alpha} f_{k,\alpha}$. 
The average current~\eqref{Jave} can then be written as $J_{\nu}(t) = \ii \Tr[\mathbf{G}_{\nu} \mathbf{C}(t)]$, where $\mathbf{G}_{\nu}$ is an anti-symmetric matrix with entries $\pm \kappa_{k,\nu}$ in the sectors connecting $\hat{c}_{p_\nu}$ and $\hat{a}_{k,\nu}$~\cite{SM}.

The noise can be obtained using the method shown in the Ref.~\cite{SM}. 
It consists of writing the noise over any interval $[t_1,t_2]$ as $D_{\nu}(t_2,t_1) = 2 \Tr[\mathbf{G}_{\nu} \mathbf{\tilde{C}}(t_2,t_1)]$, where $\mathbf{\tilde{C}}(t_2,t_1)$ is an auxiliary matrix, obtained by integrating the modified Lyapunov equation 
\begin{align}
\label{eq:auxiliary_lyapunov}
    \frac{{\rm{d}}\mathbf{\tilde{C}}(t,t_1)}{{\rm{d}}t} &= - \left[\mathbf{W}(t)\mathbf{\tilde{C}}(t,t_1) + \mathbf{\tilde{C}}(t,t_1) \mathbf{W}^\dagger(t) \right]  \\[0.2cm] \nonumber
    &-\frac{1}{2} \Big[ \mathbf{C}(t) \mathbf{G}_{\nu} [\bm{1} - \mathbf{C}(t)] + [\bm{1} - \mathbf{C}(t)] \mathbf{G}_{\nu} \mathbf{C}(t) \Big],
\end{align}
with initial condition $\mathbf{\tilde{C}}(t_1,t_1) = \bm{0}$.
The second line contains $\mathbf{C}(t)$, which is the solution of Eq.~\eqref{eq:lyapunov}, with initial condition at time $t=0$ (and not $t_1$).
Physically, we can interpret the solution
$\mathbf{\tilde{C}}(t_2,t_1)$ as turning a detector on at $t_1$ and then off at $t_2$. 
The real dynamics $\mathbf{C}(t,0)$ evolves from $t = 0$ onward, indefinitely. 
Given a window $[t_1,t_2]$, we obtain the corresponding fluctuations by integrating Eq.~\eqref{eq:auxiliary_lyapunov}. With these expressions, we can therefore analyse fluctuations over arbitrary intervals, for Hamiltonians with arbitrary time-dependence. Eqs.~\eqref{eq:lyapunov}-\eqref{eq:auxiliary_lyapunov} can be integrated using standard Runge-Kutta methods.

\emph {Time-dependent current and noise in two-terminal junctions}. We consider two metal electrodes kept at different equilibrium states and bridged by a two-site fermionic system, 
which is modulated via a
time-periodic  electric field, 
\begin{eqnarray}
\label{eq:h_s}
    \hat{H}_{\tt S}(t) = \left(\frac{eaE(t)}{2}\right) \biggl( \hat{c}^{\dagger}_1 \hat{c}_1 - \hat{c}^{\dagger}_2 \hat{c}_2 \biggr) - \Delta \left( \hat{c}^{\dagger}_1 \hat{c}_2 + \hat{c}^{\dagger}_2 \hat{c}_1 \right).
    \nonumber\\
\end{eqnarray}
Here $e$ is the electric charge, $a$ is the spacing between the two sites and $E(t) = A\cos(\omega t)$ is the electric field. 
We fix the internal coupling $\Delta$ as the energy scale of the problem. 
The two reservoirs have the same temperatures, $T_{\tt L} = T_{\tt R} = 0.1\Delta$, but a chemical potential bias $\mu_{\tt L} = 24\Delta$ and $\mu_{\tt R} = -24\Delta$~\cite{Chen:2014}. 
The spectral function of the baths are taken as $\mathcal{J}_{\tt L}(\omega) = \mathcal{J}_{\tt R}(\omega) = \Gamma,\; \forall\, \omega \in [-W, W]$, and zero otherwise, where $W$ is a cutoff energy and $\Gamma$  the effective coupling. 
We discretise each reservoir into $N$ lead-modes with energies $\varepsilon_k$ between $-W$ and $W$, such that $\gamma_{k,\alpha} = 2W / N$ and  $\kappa_{k,p} = \sqrt{\Gamma \gamma_{k,\alpha}/2\pi}$.
Throughout, we fix $\Gamma = 0.5\Delta$, $W = 100\Delta$ and $N=400$ (which sufficed to guarantee convergence of all simulations). The integration of Eqs.~\eqref{eq:lyapunov}-\eqref{eq:auxiliary_lyapunov} was carried out through fourth-order Runge-Kutta integration with a time-step $\delta t = 0.01/\Delta$.

\begin{figure}[b]
\fontsize{13}{10}\selectfont 
\centering
\includegraphics[width=1\columnwidth]{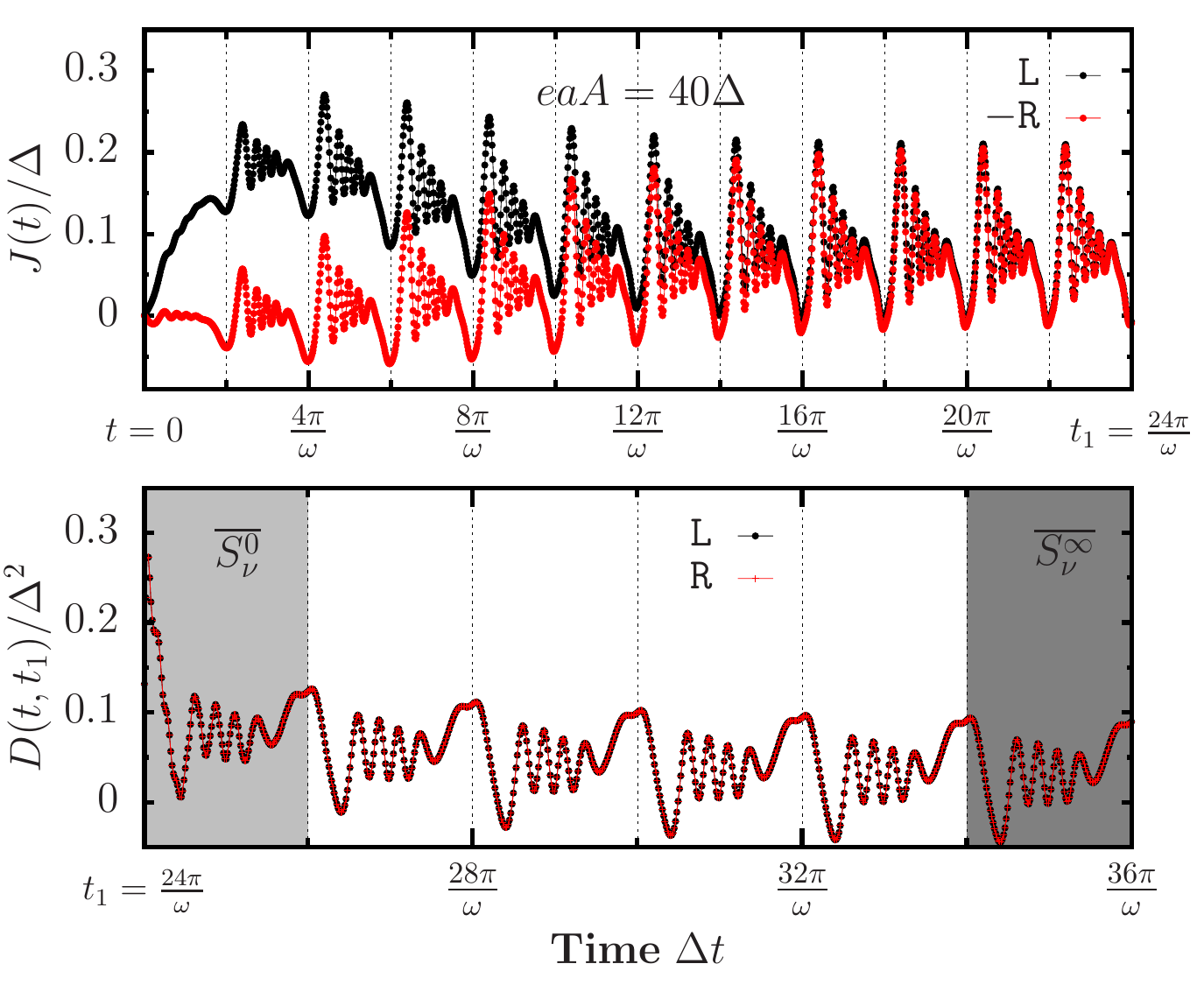}
\caption{(Top) average currents $J_{\tt L}(t)$ and $-J_{\tt R}(t)$  [Eq.~\eqref{Jave}] during multiple periods $\tau=2\pi/\omega$ of the external drive, up until the LC is reached. (Bottom)
noise $D_{\tt L/R}(t,t_1)$, starting at the LC $t_1 = 24\pi/\omega$. Integrating over the first period yields $\overline{S^{0}_{\nu}}$ in Eq.~\eqref{bar_D}.
Waiting for multiple periods and then integrating yields instead $\overline{S^{\infty}_{\nu}}$ in Eq.~\eqref{bar_S}. Parameters are described in the main text. }
\label{fig:2}
\end{figure}

Fig.~\ref{fig:2} (top) displays the instantaneous currents 
Eq.~\eqref{Jave} of the left and right reservoirs during several drive periods $\tau = 2\pi/\omega$, starting at $\mathbf{\tilde{C}}(0) = \bm{0}$, with a fixed $eaA = 40\Delta$ and $\omega = 5\Delta$. 
As can be seen, $J_{\tt L/R}$ gradually tend to the \emph{limit cycle} (LC) where $J_\nu(t+\tau) = J_\nu(t)$.  
This suggests we define the LC-averaged current as
\begin{equation}\label{bar_J}
    \overline{J}_\nu = \frac{1}{\tau} \int_{t_1}^{t_1+\tau} {\rm d} t^{\prime} J_\nu(t^{\prime}),
\end{equation}
where $t_1$ is a large enough time such that $J_\nu(t_1+\tau) = J_\nu(t_1)$.

To analyse the noise, we wait until the LC has been reached at time $t_1$, so that we eliminate any dependence on the arbitrary initial condition \footnote{Given that the Lindblad master equation is expected to be gapped, there will be a finite time until this occurs.}. 
In Fig.~\ref{fig:2} (bottom), we plot $D_{\tt L/R}(t,t_1)$, starting at $t_1 = 24\pi/\omega$. The choice of $t_1$ is arbitrary, as long as it is large enough such that $J_\nu(t_1+\tau) = J_\nu(t_1)$. Note that $D_{\tt L}(t,t_1) = D_{\tt R}(t,t_1)\;\forall t$; this is a consequence of the symmetric model parameters and it is non-generic behaviour (see Ref.~\cite{SM} for details). At $t = t_1$, we start counting particles to analyse the instantaneous noise. We find that $D_\nu(t+\tau,t_1) \neq D_\nu(t,t_1)$ over the first period [left-most grey region in Fig.~\ref{fig:2} (bottom)]. 
In fact, integrating $D_\nu(t,t_1)$ from $t_1$ to $t_1+\tau$ yields 
\begin{align}\label{bar_D}
    \overline{S^{0}_{\nu}} \defeq \frac{1}{\tau} \int_{0}^{\tau} {\rm d}t^{\prime} D_\nu(t_1+t^{\prime},t_1),
\end{align}
which is the average variance of the charge transferred over a single period after the LC. 
Similarly, integrating from $t_1$ to $t_1+2\tau$ yields the average fluctuation over two periods, and so forth, as it can also be measured. 
For $t\gg t_1,\tau$, we see in Fig.~\ref{fig:2} (bottom) that eventually the noise itself becomes periodic, and
$|D_\nu(t+\tau,t_1) - D_\nu(t,t_1)|\rightarrow 0$
for $t\rightarrow \infty$~\footnote{ 
This is a consequence of Eq.~\eqref{Ds_covariance}, which in this case can be written as 
$D(t+\tau,t_1) = D(t,t_1) + 2 \frac{{\rm{d}}}{{\rm{d}}t} \ccov\big[N(t,t_1),N(t_1,t_1-\tau)\big]$.
The last term is the rate of change of the covariance, which becomes vanishingly small when $t \gg t_1,\tau$.}. This suggests we define
\begin{equation}\label{bar_S}
    \overline{S^{\infty}_{\nu}} 
    \defeq \lim_{t\to\infty}\frac{1}{\tau} \int_{0}^{\tau} {\rm d}t' D_\nu(t+t',t_1),
\end{equation}
depicted in the right-most grey region in Fig.~\ref{fig:2} (bottom).
$\overline{S^{\infty}_{\nu}}$ is in fact the so-called \emph{LC-averaged zero-frequency 
component} of the noise~\cite{Hanggi2003}.
Despite being a more standard quantity in the context of systems with autonomous steady-states, it lacks the clear physical interpretation as $\overline{S^{0}_{\nu}}$ when time-dependent drives are present. 

\begin{figure}[t]
\fontsize{13}{10}\selectfont 
\centering
\includegraphics[width=1\columnwidth]{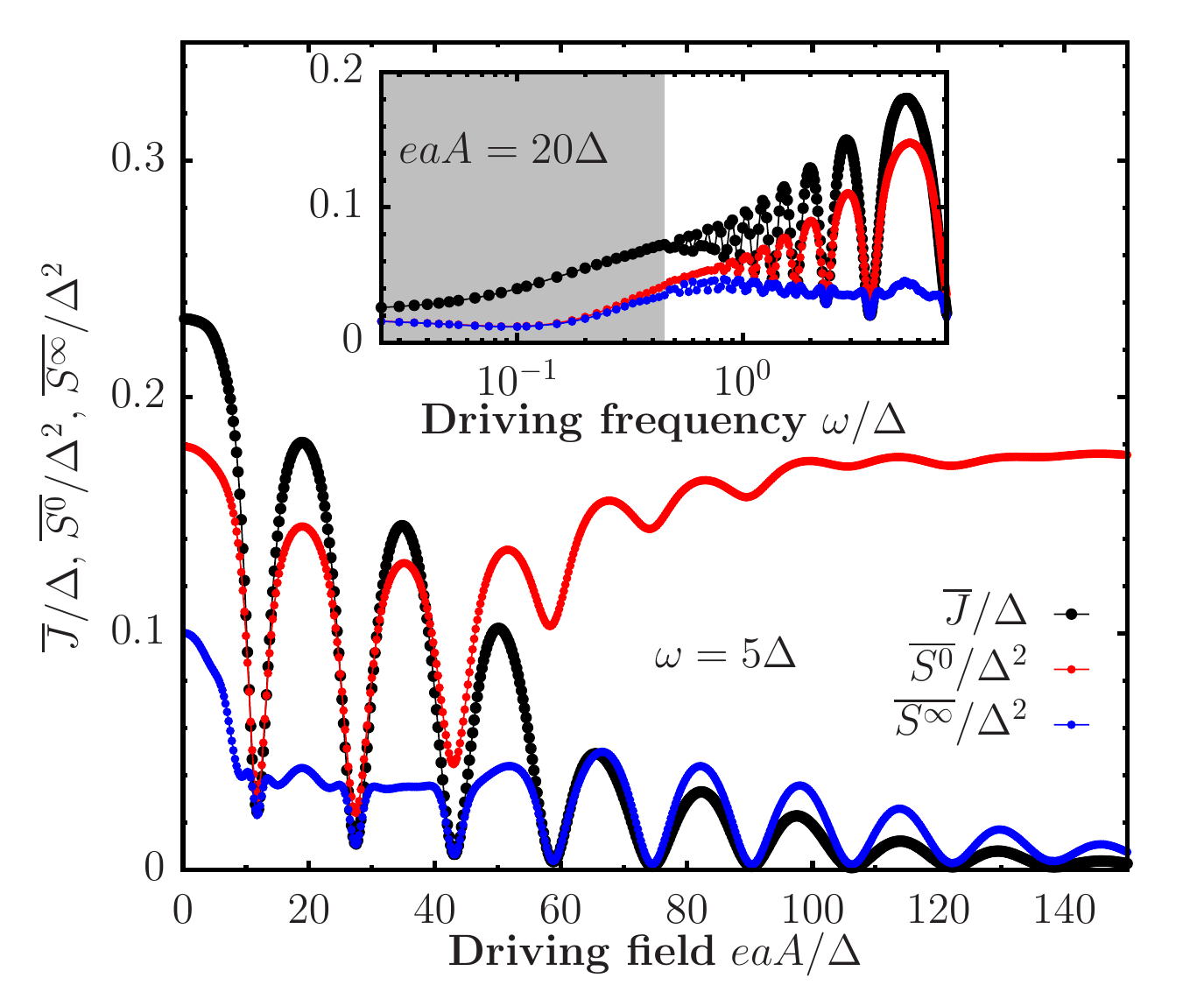}
\caption{$\overline{J}$, $\overline{S^{0}}$ and $\overline{S^{\infty}}$ [Eqs.~\eqref{bar_J}-\eqref{bar_S}] as a function of the driving field amplitude $eaA/\Delta$, in the limit cycle, with fixed frequency $\omega = 5\Delta$.
(Inset) same, but as a function of $\omega / \Delta$, with fixed $eaA = 20\Delta$. Other parameters are as in Fig.~\ref{fig:2}.}
\label{fig:3}
\end{figure}

Fig.~\ref{fig:3} displays $\overline{J}$, $\overline{S^{0}}$ and $\overline{S^{\infty}}$ as a function of the driving field strength $A$. We have suppressed the $\nu$ index, as $\tt{L}$ and $\tt{R}$ quantities are equivalent in the present case with symmetric driving and bias. This model is known to display current-suppressed minima for certain values of the driving field~\cite{Chen:2014}.
A key feature of this is that one can also systematically suppress $\overline{S^{\infty}}$~\cite{Hanggi2003}. 
In contrast, the single-period variance $\overline{S^{0}}$ displays a fundamentally different behaviour, remaining non-zero even for arbitrarily-large drive amplitudes.
This means that even though the average current is suppressed, the fluctuations of the charge exchanged within each period are not. 
This fundamental difference between $\overline{S^{\infty}}$ and $\overline{S^{0}}$ is a feature of driven systems, and depends on the frequency in question. In the inset of Fig.~\ref{fig:3} we plot $\overline{J}$, $\overline{S^{0}}$ and $\overline{S^{\infty}}$ as a function of $\omega$. We see that $\overline{S^{0}}$ and $\overline{S^{\infty}}$ coincide asymptotically in the low frequency regime, becoming identical in the non-driven case, $\omega=0$ (see Ref.~\cite{SM}). Conversely, for large frequencies, they deviate substantially.

\emph {Conclusions}. Accurate, non-perturbative computation of fluctuations of observables in driven, non-equilibrium quantum settings is a long-standing problem. Here, we have put forward a powerful, flexible method that paves the pathway into investigations of fluctuations of quantum systems out of equilibrium. 
Having access to the full dynamics of the noise we revealed that the average fluctuations contain correlations between different time periods in periodically-driven systems, arising from the non-additivity of the variance. 
Specifically our method allowed us to uncover the fundamental distinction between two measures for noise, 
$\overline{S^{0}}$ and $\overline{S^{\infty}}$. While the former is a measure for the pure variance of a physical quantity within a certain time interval, the latter contains covariance terms over time intervals [Eq.~\eqref{Ds_covariance}]. In driven systems, $\overline{S^{0}}$ thus has a clearer physical interpretation of charge fluctuations, contrasting  $\overline{S^{\infty}}$.

Future prospects include studies of light-driven materials under non-periodic modulations,
relevant to proposals for petahertz signal processing~\cite{Boolakee2022,Stockman}. In these scenarios, the full transient dynamics of currents and charge fluctuations are of the essence.

Current fluctuations can further reveal fundamental aspects of electron-electron interactions,
as demonstrated in non-driven systems \cite{Heiblum97}. By combining our FCS-mesoscopic lead framework with tensor-network techniques~\cite{Brenes:2020} one could uncover correlated-electron phenomena in nanoscale devices from the behaviour of both transient currents and their noise signals.

\emph {Acknowledgements}.
The work of M.B.~has been supported by the Centre for Quantum 
Information and Quantum Control (CQIQC) at the University of Toronto. D.S.~acknowledges support 
from NSERC and from the Canada Research Chair program. G.G.~acknowledges funding from European Union's Horizon 2020 research and innovation programme under the Marie Sk\l{}odowska-Curie Grant Agreement INTREPID, No.~101026667. A.P.~acknowledges funding from the European Union’s Horizon 2020 research and innovation programme under the Marie Sklodowska-Curie Grant Agreement No. 890884, and also from the Danish National Research Foundation through the Center of Excellence ``CCQ'' (Grant agreement no.: DNRF156). J.E.~is funded by the DFG (FOR 2724 and CRC 183) and the FQXi. Computations were performed on the Niagara supercomputer at the SciNet HPC Consortium. SciNet is funded by: the Canada Foundation for Innovation; the Government of Ontario; Ontario Research Fund - Research Excellence; and the University of Toronto. 

\bibliography{bibliography}

\clearpage
\newpage

\onecolumngrid

\setcounter{secnumdepth}{3}
\setcounter{equation}{0}
\setcounter{figure}{0}
\section*{Supplemental Material}

\renewcommand{\theequation}{S\arabic{equation}}
\renewcommand{\thesubsection}{S\arabic{section}.\arabic{subsection}}
\renewcommand{\thesection}{S\arabic{section}}
\renewcommand{\thefigure}{S\arabic{figure}}

\begin{center}
    \textbf{Particle current statistics in driven mesoscale conductors}\\
\end{center}

In this Supplemental Material we provide further details on the full counting statistics (FCS) in Sec.~\ref{sec:fcs} and on the mesoscopic leads description in Sec.~\ref{sec:therm_leads}. The full derivation on the dynamical equations of the average current and the noise for Gaussian time-dependent systems is presented in Sec.~\ref{sec:gaussian}. Finally, in Sec.~\ref{sec:noise_w} we present mathematical arguments behind the correspondence of the variance in the limit cycle $\overline{S^{0}}$ and the zero-frequency component of the noise $\overline{S^{\infty}}$ in the $\omega \to 0$ limit. We also present noise calculations in asymmetric non-equilibrium configurations, as well as different regimes of operation dictated by the temperature and the system-reservoir coupling strength. 

\section{Full counting statistics}
\label{sec:fcs}

We begin by describing the exact FCS results for measuring the charge current to the reservoirs. We consider a distinguished system captured by a Hamiltonian
$\hat{H}_{\tt S}(t)$, expressed in terms of a set of fermionic annihilation operators $\{\hat{c}_j\}$, that is coupled to $Q$ fermionic reservoirs. 
The $\alpha$-th reservoir is written in the energy basis of canonical $\{\hat{b}_{n, \alpha}\}$ operators, such that 
\begin{align}
    \hat{H}_{\tt B_\alpha} = \sum_{n=1}^{\infty} \omega_{n, \alpha} \hat{b}^{\dagger}_{n, \alpha} \hat{b}_{n, \alpha},
\end{align}
while the total reservoir Hamiltonian is given by the sum of individual terms as
\begin{align}
    \hat{H}_{\tt B} = \sum_{\alpha=1}^{Q} \hat{H}_{\tt B_\alpha}.
\end{align}
As is commonly assumed, 
the reservoirs are initially prepared in thermal states 
\begin{align}
    \hat{\rho}_{\tt B_\alpha} = e^{-\beta_{\alpha}(\hat{H}_{\tt B_\alpha}-\mu_\alpha \mathcal{N}_\alpha)} / Z_{\alpha}
\end{align}
with $\mathcal{N}_\alpha := \sum_n b_{n, \alpha}^\dagger b_{n, \alpha}$, 
$Z_{\alpha} := \Tr[e^{-\beta_{\alpha}(\hat{H}_{\tt B_\alpha}-\mu_\alpha \hat{N}_\alpha)}]$. The 
combined system composed of the distinguished system and the reservoirs configuration is thought to be prepared, initially, in a product state 
\begin{align}
\label{eq:rho_tot_init}
    \hat{\rho}_{\rm tot} = \hat{\rho}_{\tt S} \prod_{\alpha=1}^{Q} \hat{\rho}_{\tt B_\alpha},
\end{align}
reflecting what are called factorising initial conditions.
Each reservoir couples to specific system operators $\hat{c}_{p_\alpha}$ with an interaction 
\begin{align}
    \hat{H}_{\tt SB_\alpha} = \sum_{n=1}^{\infty} \lambda_{n , \alpha} \big( \hat{c}_{p_\alpha}^\dagger \hat{b}_{n, \alpha} + \hat{b}_{n, \alpha}^{\dagger} \hat{c}_{p_\alpha} \big).
\end{align}
The couplings $\lambda_{n , \alpha}$ are not necessarily weak. The corresponding reservoir spectral densities are $\mathcal{J}_\alpha(\omega) = 2\pi \sum_{n=1}^{\infty} |\lambda_{n, \alpha}|^2 \delta(\omega - \omega_{n, \alpha })$.
The total Hamiltonian for the entire configuration is given by
\begin{align}
    \hat{H}(t) = \hat{H}_{\tt S}(t) + \sum_{\alpha=1}^{Q} \left[ \hat{H}_{\tt B_\alpha} + \hat{H}_{\tt SB_\alpha} \right].
\end{align}

\subsection{Derivation of the tilted Hamiltonian}

Our interest is in the FCS for the current to a given reservoir $\nu$. This is obtained by doing a two-point measurement of $\mathcal{N}_\alpha$. 
The resulting \emph{characteristic function} (CF) is~\cite{EspositoRev2009} 
\begin{align}
\label{CF1}
    G(\chi, t, t_0) = \Tr \Big\{ U^\dagger(t, t_0) e^{\ii \chi \mathcal{N}_{\nu}} U(t, t_0) e^{-\ii \chi \mathcal{N}_{\nu}} \hat{\rho}_{\rm tot}(\chi = 0, t_0, t_0) \Big\},
\end{align}
where $\chi$ is the counting field and $U(t,t_0)= \mathcal{T} {\rm exp} 
[ -\ii \int^{t}_{t_0} {\rm d}t^{\prime} \hat{H}(t^{\prime})]$, 
with $\mathcal{T}$ being the time-ordering operator. The CF can also be written as 
\begin{align}
\label{CF2}
    G(\chi, t, t_0) = \Tr \left[ \hat{\rho}_{\rm tot}(\chi,t, t_0) \right],
\end{align}
where $\hat{\rho}_{\rm tot}(\chi,t,t_0)$ evolves according to 
\begin{align}
\label{tilted_von_neumann}
    \frac{{\rm d}\hat{\rho}_{\rm tot}(\chi,t,t_0)}{{\rm d}t} = -\ii [\hat{H}_\chi(t), \hat{\rho}_{\rm tot}(\chi,t,t_0)]_\chi,
\end{align}
where $[\hat{A}_\chi, \hat{\rho}]_\chi = \hat{A}_\chi \hat{\rho} - \hat{\rho} \hat{A}_{-\chi}$ and the newly-defined tilted Hamiltonian 
\begin{align}
\label{tilted_H}
    \hat{H}_{\chi}(t) &= \hat{H}_{\tt S}(t) + \sum_{\alpha=1}^{Q} \left[ \hat{H}_{\tt B_\alpha} + e^{\ii \chi \mathcal{N}_{\nu} / 2} \hat{H}_{\tt SB_\alpha} e^{-\ii \chi \mathcal{N}_{\nu} / 2} \right] \nonumber \\
    &= \hat{H}_{\tt S}(t) + \sum_{\alpha=1}^{Q} \hat{H}_{\tt B_\alpha} + \sum_{\alpha=1}^{Q} e^{\ii \chi  \delta_{\alpha,  \nu}\mathcal{N}_\alpha/2} \hat{H}_{\tt SB_\alpha} e^{-\ii \chi \delta_{\alpha,  \nu}\mathcal{N}_\alpha/2}.
\end{align}

Note that the total state, as well as the CF, depend on three arguments: the counting field $\chi$, the time $t$ and the time of the initial state $t_0$. Even though this notation might seem unconventional at first hand, as we shall see later, keeping track of the initial condition will be important when considering fluctuations and higher-order moments of the distribution in systems with a Hamiltonian that contains an explicit time dependence. The initial state is $\hat{\rho}_{\rm tot}(\chi = 0,t_0,t_0)$ and, in principle, it could be any system-reservoir state. However, we shall consider [as in Eq.~\eqref{eq:rho_tot_init}] initial system-reservoir product states where the reservoirs are prepared in stationary thermal states
\begin{align}
\label{eq:init_product_stationary}
\hat{\rho}_{\rm tot}(\chi = 0,t_0,t_0) = \hat{\rho}_{\tt S}(t_0, t_0) \prod_{\alpha=1}^{Q} \hat{\rho}_{\tt B_\alpha},
\end{align}
where the system-state depends on two time arguments to keep track of the initial condition and the dependence on the counting field is generated via the dynamics of Eq.~\eqref{tilted_von_neumann}.

Henceforth, for simplicity of the notation, we shall define $\hat{\rho}(\chi = 0, t, t_0) \defeq \hat{\rho}(t, t_0)$. 

It is possible to use the \emph{Baker–Campbell–Hausdorff} (BCH) formula~\cite{BreuerPetruccione} to express the counting-field term in Eq.~\eqref{tilted_H} in terms of canonical operators, which yields
\begin{align}
\label{eq:tilted_full_hamiltonian}
    \hat{H}_{\chi}(t) = \hat{H}_{\tt S}(t) + \hat{H}_{\tt B} + \sum_{\alpha=1}^{Q}\sum_{n=1}^{\infty} \lambda_{n , \alpha} \big( \hat{c}_p^\dagger \hat{b}_{n, \alpha}e^{-\ii \chi  \delta_{\alpha,  \nu} / 2} + \hat{b}_{n, \alpha}^{\dagger} \hat{c}_p e^{\ii \chi  \delta_{\alpha,  \nu} / 2} \big) .
\end{align}

At this point we would like to highlight that the CF in Eq.~\eqref{CF1} stems from the commonly-proposed scheme for the FCS, namely, the two-point measurement protocol. Such protocol has been successful in, e.g., the formulation of steady-state fluctuation relations away from equilibrium~\cite{EspositoRev2009}. However, it has been subject to recent criticism and accompanying proposals~\cite{Solinas:2015,Marti:2017,Solinas:2017,Xu:2018,Levy:2020}, particularly designed to overcome the fact that the first measurement collapses the quantum state onto an eigenstate of the initial observable. Consequently, the two-point measurement protocol fails to capture the true dynamics of general processes in which the initial state could be in coherent superposition of eigenstates. In light of Eq.~\eqref{eq:init_product_stationary}, however, we expect the two-point measurement protocol to correctly describe the statistics of the distribution since our initial state is a product state and the bath's density matrix is a statistical mixture of eigenstates.

\subsection{Formulae for the first two cumulants}
\label{subsec:formulae_cumulants}

Consider a generic tilted master equation 
\begin{align}
\frac{{\rm d}\hat{\rho}(\chi,t,t_0)}{{\rm d}t} = \mathcal{L}_{\chi}(t) \hat{\rho}(\chi,t,t_0), 
\end{align}
for some tilted Liouvillian 
that has a time dependence $t\mapsto \mathcal{L}_{\chi}(t)$ (the 
true Liouvillian being $\mathcal{L}(t)$). 
Let $N(t,t_0) = \int_{t_0}^t {\rm d}t'~I(t')$ denote the integrated current. 
If one is interested only in the first few cumulants, there is no need to consider 
the evolution of $\hat{\rho}$ under a generalised master equation [following 
Eq.~\eqref{tilted_von_neumann}]. The average current can simply be written as~(see Ref.~\cite{LandiRev2021_2})
\begin{align}
\label{fcs_average_current}
    J(t) = \frac{{\rm d}}{{\rm d}t} \langle N(t,t_0) \rangle = -\ii ~\Tr \big\{ \mathcal{L}'(t) \hat{\rho}(t,t_0)\big\},
\end{align}
where $\hat{\rho}(t,t_0)$ is the actual (non-tilted) quantum state and $\mathcal{L}'(t) = \partial_\chi \mathcal{L}_{\chi}(t) \big|_{\chi=0}$. Computing $J$ thus requires only knowledge of $\hat{\rho}(\chi = 0,t,t_0)$ and not $\hat{\rho}(\chi,t,t_0)$. 
Similarly, the instantaneous noise can be written as  
\begin{align}
\label{fcs_diffusion_coefficient}
    D(t,t_0) \defeq \frac{{\rm d}}{{\rm d}t} \Bigg(\langle N^2(t,t_0) \rangle - \langle N(t,t_0) \rangle^2\Bigg) = - \Tr\Big\{ \mathcal{L}''(t)\hat{\rho}(t,t_0)\Big\} - 2 \Tr \Big\{ \mathcal{L}'(t) \hat{\sigma}(t,t_0) \Big\}, 
\end{align}
where $\hat{\sigma}(t,t_0)$ is an auxiliary variable that satisfies the differential equation (see Sec.~II.I.3 in Ref.~\cite{LandiRev2021_2})
\begin{align}
\label{fcs_sigma_equation}
    \frac{{\rm d}\hat{\sigma}(t,t_0)}{{\rm d}t} = \mathcal{L}(t)\hat{\sigma}(t,t_0) + \Big[ \mathcal{L}'(t)\hat{\rho}(t,t_0) - \hat{\rho}(t,t_0) \Tr[\mathcal{L}'(t) \hat{\rho}(t,t_0)] \Big], 
\end{align}
with initial condition $\hat{\sigma}(t_0,t_0) = 0$. 
Thus, to compute $D(t,t_0)$ we only require the dynamics of $\hat{\rho}(t,t_0)$ and $\hat{\sigma}(t,t_0)$. 

\section{Mesoscopic leads transformation}
\label{sec:therm_leads}

Next we implement a unitary transformation to the \emph{mesoscopic leads configuration} [Fig.~1 (main text)]. 
The basic idea is that, instead of having the system directly coupled to an infinite number of modes $\{\hat{b}_{n, \alpha}\}$ for the $\alpha$-th reservoir, we couple it instead to a finite number $N_{\alpha}$ of \emph{lead modes} $\{\hat{a}_\alpha\}$, where each of the $\{\hat{a}_\alpha\}$ modes are coupled to their own Markovian environment $\tt E$, described by modes $\{d_{j,\alpha}\}$. This can be seen as a variant of a mappings of open quantum systems onto chain representations and Markovian embeddings (see, e.g., Ref.~\cite{Woods} and references therein).  The key difference, however, is that not many direct couplings are mapped to single chains with modulated couplings, but instead to collections of single fermionic degrees of freedom, each of which is coupled to its own Markovian environment. This is a highly desirable feature for the purposes of this work. Specifically, the transformation is implemented in such a way that the system-bath coupling is transformed into a system-lead as
\begin{equation}
    \hat{H}_{\tt SB_\alpha} \to \hat{H}_{\tt SL_{\alpha}} = \sum_{k=1}^{N_\alpha} \kappa_{k, \alpha} \big( \hat{c}_p^\dagger \hat{a}_{k, \alpha} + \hat{a}_{k, \alpha}^\dagger \hat{c}_p\big).
\end{equation}
The bath Hamiltonian, in contrast, transforms as 
\begin{align}
\label{eq:bath_total_ham}
    \hat{H}_{\tt B_\alpha} &\mapsto \hat{H}_{{\tt L}_{\alpha}} + \hat{H}_{{\tt E}_{\alpha}} + \hat{H}_{{\tt LE}_{\alpha}} = \sum_{k=1}^{N_{\alpha}} \varepsilon_{k , \alpha}   \hat{a}^{\dagger}_{k, \alpha} \hat{a}_{k, \alpha} + \sum_{k=1}^{N_{\alpha}} \sum_{j=1}^{\infty} \Omega_{j,k,\alpha} \hat{d}^{\dagger}_{j,k,\alpha} \hat{d}_{j,k,\alpha} + \sum_{k=1}^{N_{\alpha}} \sum_{j=1}^{\infty} \Lambda_{j,k,\alpha} \big( \hat{a}_{k, \alpha}^\dagger \hat{d}_{j,k,\alpha} + \hat{d}_{j,k,\alpha}^\dagger \hat{a}_{k, \alpha}\big) .
\end{align}
Hence, the total Hamiltonian is mapped as 
\begin{align}
    \hat{H}(t) &\mapsto \hat{H}_{\tt S}(t) + \sum_{\alpha=1}^{Q} \left[ \hat{H}_{{\tt L}_{\alpha}} + \hat{H}_{{\tt E}_{\alpha}} + \hat{H}_{{\tt LE}_{\alpha}} + \hat{H}_{{\tt SL}_{\alpha}}  \right] \nonumber \\
    &\defeq \hat{H}_{\tt S}(t) + \hat{H}_{\tt L} + \hat{H}_{\tt E} + \hat{H}_{\tt LE} + \hat{H}_{\tt SL}.
\end{align}
At first glance, this extension would seem to only convolute the problem further. However, if we assume that each single energy mode $\varepsilon_{k , \alpha}$ is coupled to an effective infinite reservoir with a flat spectral density
\begin{align}
\label{eq:flat_spectral_density}
    \mathcal{J}_{k, \alpha}(\omega) = \sum_{j=1}^{\infty} |\Lambda_{j,k,\alpha}|^2 \delta(\omega - \Omega_{j,k,\alpha}) \defeqr \gamma_{k, \alpha},
\end{align}
the resulting effective spectral density to the system of interest with the addition of the single mode will take a Lorentzian form~\cite{Brenes:2020}. The key insight in the mesoscopic leads transformation~\cite{Imamoglu1994,Garraway1997a,Garraway1997b,Subotnik:2009,Brenes:2020}  then lies in the realisation that any continuous spectral function may be discretised into $k$ Lorentzian functions, each centred at energy $\varepsilon_{k , \alpha}$ of width $\gamma_{k , \alpha}$ for the $\alpha$-th reservoir. We name this the \emph{effective spectral function} with form
\begin{align}
\label{eq:effective_spectral_density}
    \mathcal{J}^{\rm eff}_{\alpha}(\omega) = \sum_{k=1}^{N_{\alpha}} \frac{|\kappa_{k, \alpha}|^2 \gamma_{k , \alpha}}{(\omega - \varepsilon_{k , \alpha}) + (\gamma_{k, \alpha}/2)^2},
\end{align}
which converges to the true spectral density, $\mathcal{J}_{\alpha}^{\rm eff}(\omega) \mapsto \mathcal{J}_{\alpha}(\omega)$, in the limit of $N_{\alpha} \to \infty$
\cite{Brenes:2020}.
Furthermore, the $\kappa_{k, \alpha}$ are fixed by the true spectral function via
\begin{align}
    \kappa_{k, \alpha}^2 = \frac{\mathcal{J_{\alpha}}(\varepsilon_{k , \alpha}) \gamma_{k, \alpha}}{2\pi},
\end{align}
where we have assumed the dissipative couplings to be $\gamma_{k, \alpha} = \varepsilon_{(k+1)\alpha} - \varepsilon_{k, \alpha}$ and each $\varepsilon_{k, \alpha}$ are sampled equidistantly in the energy space of the bandwidth of the $\alpha$-th reservoir. As $N_{\alpha}$ increases, the $\gamma_{k, \alpha}$ will decrease asymptotically which allows one to justify a Lindblad master equation for the dynamics of each mode coupled to the system~\cite{Brenes:2020}.
We define $\hat{\rho} \defeq \hat{\rho}_{\tt SL}$ as the quantum state of the joint system composed of the distinguished system and the leads, obtained after tracing over the residual $\tt E$ bath. 
In the specific case without the counting fields, the dynamics is found to be Markovian and the time dependent quantum state satisfies the Lindblad master equation (see Sec.~\ref{sec:generalised_master_eq})
\begin{equation}
\label{eq:M}
    \frac{{\rm d}\hat{\rho}(t,t_0)}{{\rm d}t} = -\ii [\hat{H}_{\tt S}(t) + \hat{H}_{\tt L} + \hat{H}_{\tt SL}, \hat{\rho}(t,t_0)] + \sum_{\alpha=1}^{Q} \sum_{k=1}^{N_{\alpha}} \Big\{ \gamma_{k, \alpha} (1-f_{k, \alpha}) D[\hat{a}_{k, \alpha} ] + \gamma_{k, \alpha} f_{k, \alpha} D[\hat{a}_{k, \alpha}^\dagger]\Big\},
\end{equation}
where $f_{k, \alpha} \defeq f(\varepsilon_{k, \alpha})$ is the Fermi-Dirac distribution evaluated at the energy of the lead mode, and $D[\hat{L}] \defeq \hat{L} \hat{\rho} \hat{L}^\dagger - \frac{1}{2}\{\hat{L}^\dagger \hat{L}, \hat{\rho}\}$. The average particle current can be computed directly in this case~\cite{Brenes:2020}. By definition 
\begin{align}
    J_{\nu}(t) = \frac{{\rm d} \langle \mathcal{N}_{\nu} \rangle}{{\rm d}t} = \ii \langle [\hat{H}_{\tt SB}, \mathcal{N}_{\nu}] \rangle. 
\end{align}
Since $\hat{H}_{\tt SB} = \hat{H}_{\tt SL}$ in the leads picture, we are then left only with 
\begin{align}
\label{average_particle_current}
    J_{\nu}(t) = \ii \sum_{k=1}^{N_{\nu}} \kappa_{k, \nu} \langle \hat{c}^\dagger_{p_\nu} \hat{a}_{k, \nu} - \hat{a}_{k, \nu}^\dagger \hat{c}_{p_\nu} \rangle. 
\end{align}

\subsection{Tilted Hamiltonian in the mesoscopic leads transformation}

Going back to Eq.~\eqref{eq:tilted_full_hamiltonian}, only the $\hat{H}_{\tt SB_{\alpha}}$ gets tilted with the counting field $\chi$. Through the BCH formula we may pull the counting fields out of the sum, and write
\begin{align}
    \hat{H}^{\chi}_{\tt SB_{\nu}} &= e^{\ii \chi  \mathcal{N}_\nu/2}  \hat{H}_{\tt SB_\nu} e^{-\ii \chi \mathcal{N}_\nu/2} \nonumber \\[0.2cm]
    &= e^{-\ii \chi / 2} \sum_{n=1}^{\infty} \lambda_{n , \nu} \hat{c}_p^\dagger \hat{b}_{n, \nu} + e^{\ii \chi / 2} \sum_{n=1}^{\infty} \lambda_{n , \nu} \hat{b}_{n, \nu}^{\dagger} \hat{c}_p,
    \label{SM_tmp1231231}
\end{align}
which is exactly the form to which $\hat{H}_{\tt SL_{\nu}}$ gets mapped in the mesoscopic leads transformation. That is,
\begin{align}
\label{eq:tilted_interaction_hamiltonian_leads}
     \hat{H}^{\chi}_{\tt SB_{\nu}} &\mapsto \hat{H}^{\chi}_{\tt SL_{\nu}} \nonumber \\
     & = \sum_{k=1}^{N_\nu} \kappa_{k, \nu} \big( e^{-\ii \chi / 2} \hat{c}_p^\dagger \hat{a}_{k, \nu} +  e^{\ii \chi / 2} \hat{a}_{k, \nu}^\dagger \hat{c}_p\big).
\end{align}
Eq.~\eqref{SM_tmp1231231} is the key technical result of this calculation. It shows that counting the change in particles in the bath is  equivalent to putting a counting field in the lead modes. The full tilted Hamiltonian in the mesoscopic leads transformation, then, reads
\begin{align}
\label{eq:tilted_total_hamiltonian_leads}
    \hat{H}_{\chi}(t) &= \hat{H}_{\tt S}(t) + \hat{H}_{\tt L} + \hat{H}_{\tt E} + \hat{H}_{\tt LE} + \hat{H}^{\chi}_{\tt SL} \nonumber \\
    &=\hat{H}_{\tt S}(t) + \hat{H}_{\tt L} + \hat{H}_{\tt E} + \hat{H}_{\tt LE} + \sum_{\alpha=1}^{Q} \sum_{k=1}^{N_\alpha} \kappa_{k, \alpha} \big( e^{-\ii \chi \delta_{\alpha,  \nu} / 2} \hat{c}_p^\dagger \hat{a}_{k, \alpha} +  e^{\ii \chi \delta_{\alpha,  \nu} / 2} \hat{a}_{k, \alpha}^\dagger \hat{c}_p\big).
\end{align}

\subsection{Generalised master equation}
\label{sec:generalised_master_eq}

We return to the computation of the characteristic function in Eq.~\eqref{CF2}. 
Our starting point is Eq.~\eqref{tilted_von_neumann} and we aim to trace over the residual environment $\tt E$, to obtain a generalised master equation for $\hat{\rho}(\chi,t,t_0) \defeq \hat{\rho}_{\tt SL}(\chi,t,t_0) = \Tr_{\tt E}[ \hat{\rho}_{\tt SLE}(\chi,t,t_0) ]$. We can then write Eq.~\eqref{CF2} as $G(\chi,t,t_0) = \Tr_{\tt SL} \hat{\rho}(\chi,t,t_0)$.
To be consistent with Eq.~\eqref{eq:M}, we want to derive a \emph{local} generalised master equation, acting only on the lead modes. We therefore move to the interaction picture with respect only to $\hat{H}_0 =  \hat{H}_{\tt L} + \hat{H}_{\tt E}$. This yields 
\begin{align}
\label{eq:total_evo_rho}
    \frac{{\rm d}\hat{\rho}_{\tt SLE}(\chi,t,t_0)}{{\rm d}t} = -\ii [\hat{H}_{\tt S}(t) + \hat{H}^{\chi}_{\tt SL}(t) + \hat{H}_{\tt LE}(t), \hat{\rho}_{\tt SLE}(\chi,t,t_0)]_\chi. 
\end{align}
The formal solution is 
\begin{align}
    \hat{\rho}_{\tt SLE}(\chi,t,t_0) = \hat{\rho}_{\tt SLE}(\chi,t_0,t_0) - \ii \int_{t_0}^t {\rm d}t' [\hat{H}_{\tt S}(t') + \hat{H}^{\chi}_{\tt SL}(t') + \hat{H}_{\tt LE}(t'), \hat{\rho}_{\tt SLE}(\chi, t',t_0)]_\chi.
\end{align}
As usual~\cite{BreuerPetruccione}, we reinsert this in Eq.~\eqref{eq:total_evo_rho}.
However, to obtain a local master equation, we only reinsert it in the terms containing support on $\tt E$~\cite{LandiRev2021}. That is, we write, after tracing over $\tt E$
\begin{equation}
    \frac{{\rm d}\hat{\rho}(\chi,t,t_0)}{{\rm d}t} = -\ii [\hat{H}_{\tt S}(t) + \hat{H}_{\tt SL}^\chi(t), \hat{\rho}(\chi,t,t_0)]_\chi
    -\int_{t_0}^t {\rm d}t'~\Tr_{\tt E}[\hat{H}_{\tt LE}(t),[\hat{H}_{\tt LE}(t'), \hat{\rho}_{\tt SLE}(\chi,t',t_0)]_\chi]_\chi.
\end{equation}
Here, we have already neglected terms containing combinations of $\hat{H}_{\tt S}(t) + \hat{H}_{\tt SL}^\chi(t)$ with $\hat{H}_{\tt LE}(t')$, which vanish since they involve expectation values of linear operators with support on $\tt E$. Applying the usual Born-Markov~\cite{BreuerPetruccione} approximations then leads to 
\begin{align}
\label{gmeq_start}
    \frac{{\rm d}\hat{\rho}(\chi,t,t_0)}{{\rm d}t} &= -\ii [\hat{H}_{\tt S}(t) + \hat{H}_{\tt SL}^\chi(t), \hat{\rho}(\chi,t,t_0)]_\chi
    -\int_{t_0}^\infty {\rm d}t'~\Tr_{\tt E}[\hat{H}_{\tt LE}(t),[\hat{H}_{\tt LE}(t-t'),\hat{\rho}(\chi,t,t_0)\hat{\rho}_{\tt E}]_\chi]_\chi \nonumber \\
    & = -\ii [\hat{H}_{\tt S}(t) + \hat{H}_{\tt SL}^\chi(t), \hat{\rho}(\chi,t,t_0)]_\chi
    -\int_{t_0}^\infty {\rm d}t'~\Tr_{\tt E}[\hat{H}_{\tt LE}(t),[\hat{H}_{\tt LE}(t-t'),\hat{\rho}(\chi,t,t_0)\hat{\rho}_{\tt E}]],
\end{align}
where we have suppressed the subscript $\chi$ in the commutators appearing inside the integral, since in the mesoscopic transformation the $\hat{H}_{\tt LE}$ interaction Hamiltonian no longer depends on the counting field. Consequently, the integral on the right-hand side now becomes the standard integral appearing in the derivation of Born-Markov-Secular master equations~\cite{BreuerPetruccione,LandiRev2021}, with the exception that the trace over $\tt E$ now refers to tracing out environment degrees of freedom from each fermionic site composing the lead of the $\alpha$-th reservoir.

Note that the lead-environment interaction is of the usual form~\cite{LandiRev2021_2} $\hat{H}_{\tt LE_{\alpha}}(t) = \sum_{\beta} \hat{A}_{\beta}(t) \hat{B}_{\beta}(t)$, where $\hat{A}_{\beta}(t)$ and $\hat{B}_{\beta}(t)$ are operators with support over lead and environment degrees of freedom, respectively. After expanding the commutators inside the integral in Eq.~\eqref{gmeq_start} and tracing over $\tt E$, we obtain
\begin{align}
\frac{{\rm d}\hat{\rho}(\chi,t,t_0)}{{\rm d}t} = -\ii [\hat{H}_{\tt S}(t) + \hat{H}_{\tt SL}^\chi(t), \hat{\rho}(\chi,t,t_0)]_\chi - \sum_{\beta \gamma} \int_{t_0}^\infty {\rm d}t'~\{ &C_{\beta \gamma}(t') [\hat{A}_{\beta}(t), \hat{A}_{\gamma}(t - t')\hat{\rho}(\chi,t,t_0)] \nonumber \\
&+ C_{\gamma \beta}(-t') [\hat{\rho}(\chi,t,t_0)\hat{A}_{\gamma}(t - t'), \hat{A}_{\beta}(t)] \}.
\end{align}
Note that due to the bilinear form of $\hat{H}_{\tt LE_{\alpha}}(t)$, it follows that $\beta, \gamma = 1,2$. Above, we have defined the environment correlations
\begin{align}
C_{\beta \gamma}(t_1,t_2) = \Tr_{\tt E}[\hat{B}_{\beta}(t_1)\hat{B}_{\gamma}(t_2) \hat{\rho}_{\tt E}] = \Tr_{\tt E}[\hat{B}_{\beta}(t_1 - t_2)\hat{B}_{\gamma} \hat{\rho}_{\tt E}] \defeq C_{\beta \gamma}(t_1 - t_2),
\end{align}
where we have assumed that $\hat{\rho}_{\tt E}$ is a stationary (thermal) state, $[\hat{H}_{\tt E}, \hat{\rho}_{\tt E}] = 0$. Consequently, only cross-correlations are non-zero ($\beta \neq \gamma$) and will lead to expressions which depend on the Fermi-Dirac distribution. We also assume that thermal correlations between environment operators acting on different lead modes vanish. 

Note that the operators $\hat{A}_{\beta}(t)$ are just canonical creation and annihilation operators with support over lead modes, which evolve according to $\hat{a}^{\dagger}_{k,\alpha}(t) = \hat{a}^{\dagger}_{k,\alpha} e^{\ii \varepsilon_{k, \alpha} t}$ and $\hat{a}_{k,\alpha}(t) = \hat{a}_{k,\alpha} e^{-\ii \varepsilon_{k, \alpha} t}$. These expressions allow one to evaluate the integral in Fourier space using our defined spectral density in Eq.~\eqref{eq:flat_spectral_density}. We remark that only the residual environments are defined by a flat spectral function, while the effective spectral function can have any continuous form according to Eq.~\eqref{eq:effective_spectral_density}.

It is then straightforward, albeit somewhat cumbersome, to then arrive to the final form of the generalised master equation in the Schr\"odinger picture
\begin{align}
\label{eq:generalised_master_equation_leads}
    \frac{{\rm d}\hat{\rho}(\chi,t,t_0)}{{\rm d}t} = -\ii [\hat{H}_{\tt S}(t) + \hat{H}_{\tt L} + \hat{H}_{\tt SL}^\chi, \hat{\rho}(\chi,t,t_0)]_\chi
    + \sum_{\alpha=1}^{Q} \sum_{k=1}^{N_{\alpha}} \Big\{ \gamma_{k, \alpha} (1-f_{k, \alpha}) D[\hat{a}_{k, \alpha} ] + \gamma_{k, \alpha} f_{k, \alpha} D[\hat{a}_{k, \alpha}^\dagger]\Big\},
\end{align}
with $\hat{H}_{\tt SL}^\chi$ defined in Eq.~\eqref{eq:tilted_total_hamiltonian_leads}. It is interesting to note that the usual Lamb-shift Hamiltonian does not appear here, since by construction the environments that couple to each lead mode have flat spectral densities~\footnote{In particular, the principal value integrals vanish due to the flat spectral function.}.

\subsection{Average current and noise}

We thus have an equation of the form
\begin{align}
    \frac{{\rm d}\hat{\rho}(\chi,t,t_0)}{{\rm d}t} = \mathcal{L}_{\chi}(t) \hat{\rho}(\chi,t,t_0)
\end{align}
with $\mathcal{L}_{\chi}(t)$ defined from Eq.~\eqref{eq:generalised_master_equation_leads}. Note that the explicit time dependence of $\mathcal{L}_{\chi}(t)$ comes from the time dependence in $\hat{H}_{\tt S}(t)$. This expression allows us to obtain expressions for the first two cumulants in terms of canonical operators from Eqs.~\eqref{fcs_average_current}-\eqref{fcs_sigma_equation}, which depend on $\mathcal{L}' = \partial_\chi \mathcal{L}_{\chi}(t) \big|_{\chi=0}$ and $\mathcal{L}'' = \partial^2_\chi \mathcal{L}_{\chi}(t) \big|_{\chi=0}$. We remove the time dependence on both $\mathcal{L}'$ and $\mathcal{L}''$, since these terms will not contain the explicit time dependence that is only present in $\hat{H}_{\tt S}(t)$, which in turn does not depend on $\chi$ and, therefore, $\mathcal{L}'$ and $\mathcal{L}''$ are static. From Eq.~\eqref{eq:tilted_total_hamiltonian_leads} and Eq.~\eqref{eq:generalised_master_equation_leads}, we have
\begin{align}
    \mathcal{L}'\rho &= - \frac{1}{2} \sum_{k=1}^{N_{\nu}} \kappa_{k, \nu}\big\{ \hat{c}^\dagger_p \hat{a}_{k, \nu} - \hat{a}_{k, \nu}^\dagger \hat{c}_p, \rho\big\}
\end{align}
and
\begin{align}
    \mathcal{L}''\rho &= \frac{\ii}{4} \sum_{k=1}^{N_{\nu}} \kappa_{k, \nu} \big[ \hat{c}^\dagger_p \hat{a}_{k, \nu} + \hat{a}_{k, \nu}^\dagger \hat{c}_p,\rho\big].
\end{align}
The equations for the instantaneous current and noise follow from Eqs.~\eqref{fcs_average_current}-\eqref{fcs_sigma_equation}. The instantaneous current
\begin{align}
    J_{\nu}(t) = \ii \sum_{k=1}^{N_{\nu}} \kappa_{k, \nu} \langle \hat{c}^\dagger_p \hat{a}_{k, \nu} - \hat{a}_{k, \nu}^\dagger \hat{c}_p \rangle,
\end{align}
with the expectation value taken over $\hat{\rho}(t,t_0)$.
For the noise, the term proportional to $\mathcal{L}''$ in Eq.~\eqref{fcs_diffusion_coefficient} vanishes since it is a commutator. We are thus left with 
\begin{align}
    D_{\nu}(t,t_0) = 2 \sum_{k=1}^{N_{\nu}} \kappa_{k, \nu} \Tr \{ \big( \hat{c}^\dagger_p \hat{a}_{k, \nu} - \hat{a}_{k, \nu}^\dagger \hat{c}_p \big) \hat{\sigma}(t,t_0) \},
\end{align}
where $\hat{\sigma}(t,t_0)$ is the solution to the time-dependent differential equation
\begin{align}
    \frac{{\rm d}\hat{\sigma}(t,t_0)}{{\rm d}t} = \mathcal{L}(t)\hat{\sigma}(t,t_0) - \sum_{k=1}^{N_{\nu}} \kappa_{k, \nu} \Big[ \big( \hat{c}^\dagger_p \hat{a}_{k, \nu} - \hat{a}_{k, \nu}^\dagger \hat{c}_p \big)\hat{\rho}(t,t_0) - \langle \hat{c}^\dagger_p \hat{a}_{k, \nu} - \hat{a}_{k, \nu}^\dagger \hat{c}_p \rangle \hat{\rho}(t,t_0)  \Big],
\end{align}
with the initial condition $\hat{\sigma}(t_0,t_0) = 0$ and $\mathcal{L}(t)$ the actual, non-tilted Liouvillian defined in Eq.~\eqref{eq:M}.

\section{Gaussian systems}
\label{sec:gaussian}

We now specialise our results to the situation where the system Hamiltonian $\hat{H}_{\tt S}(t)$ is quadratic in the fermionic operators with a generic time-dependence which needs not be periodic.
For simplicity, we focus here on the problem of a fermionic system of $L$ sites coupled to a {\em single} fermionic reservoir, i.e., $Q = 1$. The generalisation to multiple reservoirs, as we shall see, will follow straightforwardly. 
Thus, overall there will be $L + N$ modes, for which we shall use a set of $\hat{b}_j$ operators for the entire system plus lead configuration for simplicity, $j = 1,\cdots,L,L+1,\cdots,L+N$.
The non-tilted master equation reads 
\begin{equation}
    \frac{{\rm d}\hat{\rho}_{\tt SL}(t,t_0)}{{\rm d}t} = \mathcal{L}(t)\hat{\rho}_{\tt SL}(t,t_0) = -\ii [\hat{H}_{\tt S}(t) + \hat{H}_{\tt L} + \hat{H}_{\tt SL}, \hat{\rho}_{\tt SL}(t,t_0)] + \sum_{k=1}^{N} \Big\{ \gamma_{k} (1-f_{k}) D[\hat{a}_{k} ] + \gamma_{k} f_{k} D[\hat{a}_{k}^\dagger]\Big\}.
\end{equation}
The Gaussian nature of the problem means that we may write
\begin{equation}
    \hat{H}(t) = H_{\tt S}(t) + \hat{H}_{\tt L} + \hat{H}_{\tt SL} = \sum_{i,j=1}^{L+N} h_{i, j}(t) \hat{b}_i^\dagger \hat{b}_j,
\end{equation}
where only the terms $i,j \in [1,L]$ contain the explicit time-dependence. In principle, there exists no need to express the degrees of freedom of $\hat{H}_{\tt S}(t)$ in either configurational or energy space, as long as the basis of $\hat{b}_j$ is consistent. For instance, for the case of a single reservoir $Q=1$ coupled to the first site $p = 1$ of a system with $L$ fermionic sites, the matrix elements $[\mathbf{H}(t)]_{i, j} = h_{i, j}(t)$ may be displayed in matrix form as
\begin{equation}
\label{eq:full_h_mat}
    \mathbf{H}(t) = \begin{pmatrix}
    h_{1,1}(t) & \ldots & h_{1,L}(t) & \kappa_1 & \ldots & \kappa_N \\[0.2cm]
    \vdots & \ddots & \vdots & 0 & \ldots & 0 
    \\[0.2cm]
    h_{L,1}(t) & \ldots & h_{L,L}(t) & 0 & \ldots & 0 
    \\[0.2cm]
    \kappa_1 & 0 & 0 & \varepsilon_1 & \ldots & 0
    \\[0.2cm]
    \vdots & \vdots & \vdots & \vdots & \ddots & 0
    \\[0.2cm]
    \kappa_N & 0 & 0 & 0 & 0 & \varepsilon_N
    \end{pmatrix},
\end{equation}
where the first $i,j \in [1,L]$ elements are written in configuration space. We can now define a covariance matrix $\mathbf{C}$ which is positive semi-definite $\mathbf{C}\geq 0$ by construction and that has entries $[\mathbf{C}]_{i, j} = \textrm{Tr}[\hat{\rho} \hat{b}_j^\dagger \hat{b}_i]$.
One may then verify that its time evolution is governed by
\begin{align}
\label{eq:lyapunovsm}
    \frac{{\rm d}\mathbf{C}(t)}{{\rm d}t} =  -\left(\mathbf{W}\mathbf{C} + \mathbf{C}\mathbf{W}^{\dagger} \right) + \mathbf{F},
\end{align}
where 
\begin{equation}
    \mathbf{W} = \ii \mathbf{H} + \frac{\bm{\gamma}}{2}, 
    \qquad 
    [\mathbf{F}]_{k,k} = F_k = \gamma_k f_k,
\end{equation}
with $\bm{\gamma}$ a diagonal matrix with entries $[\bm{\gamma}]_{k,k} =: \gamma_k$ for $k = L+1,\cdots,L+N$ and zero otherwise; similarly for $\mathbf{F}$. For multiple-reservoir configurations, the matrix is extended accordingly and the $\kappa_{k, \alpha}$ couplings will be located in the row/column corresponding to the fermionic system-site index to which the reservoir is coupled, while the self-energies $\varepsilon_{k, \alpha}$ will remain in the diagonals of each diagonal block, as in Eq.~\eqref{eq:full_h_mat}.
Eq.~\eqref{eq:lyapunovsm} gives a closed-form expression for the dynamics of the correlation matrix in Gaussian systems. 

We shall now go back to Eqs.~\eqref{fcs_average_current}-\eqref{fcs_diffusion_coefficient} to evaluate the first two moments of the charge distribution in the Gaussian case, which can be explicitly computed. 
The generalised Liouville operator reads
\begin{align}
    \mathcal{L}_{\chi} = -\ii [\hat{H}_{\tt S}(t) + \hat{H}_{\tt L} + \hat{H}_{\tt SL}^{\chi}, \bullet] + \sum_{k=L+1}^{L+N} \Big\{ \gamma_{k} (1-f_{k}) D[\hat{b}_{k} ] + \gamma_{k} f_{k} D[\hat{b}_{k}^\dagger]\Big\},
\end{align}
where $\hat{H}^{\chi}_{\tt SL} = \sum_{k=1}^{N} \kappa_{k} \big( e^{-\ii \chi / 2} \hat{c}_p^\dagger \hat{a}_{k} +  e^{\ii \chi / 2} \hat{a}_{k}^\dagger \hat{c}_p\big)$ as in Eq.~\eqref{eq:tilted_interaction_hamiltonian_leads} and this is the only $\chi$-dependent term. In our simplified notation, for the reservoir coupled to the first fermionic system-site, we find
\begin{align}
    \hat{H}^{\chi}_{\tt SL} = \sum_{k=L+1}^{L+N} \kappa_{k} \big( e^{-\ii \chi / 2} \hat{b}_1^\dagger \hat{b}_{k} +  e^{\ii \chi / 2} \hat{b}_{k}^\dagger \hat{b}_1\big),
\end{align}
then
\begin{align}
    \mathcal{L}' \defeq \left[ \frac{\partial \mathcal{L}_{\chi}}{\partial \chi} \right]_{\chi = 0} = -\frac{1}{2} \sum_{k=L+1}^{L+N} \kappa_{k} \{ \hat{b}_1^\dagger \hat{b}_{k} - \hat{b}_{k}^\dagger \hat{b}_1, \bullet \}.
\end{align}
Furthermore, from this structure we may define a matrix $\mathbf{G}$, with entries
\begin{align}
\label{eq:g_elements}   [\mathbf{G}]_{1,k} = -[\mathbf{G}]_{k,1} = \kappa_k, 
\end{align}
where $k = L+1, \cdots, L+N$, and zero otherwise. Then
\begin{align}
    \mathcal{L}' = -\frac{1}{2}\sum_{i, j} [\mathbf{G}]_{i, j} \{ \hat{b}^{\dagger}_i \hat{b}_j, \bullet \} .
\end{align}
From Eq.~\eqref{fcs_average_current}, we then have
\begin{align}
    J(t) = -\ii \Tr[\mathcal{L}' \hat{\rho}(t,t_0)] = \ii \Tr [\mathbf{G} \mathbf{C}(t)].
\end{align}
For the  noise, we get instead
\begin{align}
    D(t,t_0) = 2 \sum_{k=L+1}^{L+N} \kappa_k \Tr \left[ (\hat{b}_1^\dagger \hat{b}_{k} - \hat{b}_{k}^\dagger \hat{b}_1) \hat{\sigma}(t) \right].
\end{align}
Similarly as before, we may define an auxiliary covariance matrix
$\tilde{\mathbf{C}}(t,t_0)$ with entries
\begin{align}
    [\tilde{\mathbf{C}}(t,t_0)]_{i, j} \defeq 
    \Tr [\hat{b}^{\dagger}_j \hat{b}_i \hat{\sigma}(t,t_0)],
\end{align}
such that
\begin{align}
    D(t,t_0) = 2\Tr[\mathbf{G}\tilde{\mathbf{C}}(t,t_0)].
\end{align}
All that is left to be identified is the equation that leads the dynamics of $\tilde{\mathbf{C}}(t,t_0)$. For this purpose, we go back to Eq.~\eqref{fcs_sigma_equation} and write, in terms of the auxiliary covariance matrix,
\begin{align}
\label{C_tilde_basic_equation_particle_current}
    \frac{{\rm d} \tilde{C}_{i, j}(t,t_0)}{{\rm d}t} = - \left[ \mathbf{W}(t) \tilde{\mathbf{C}}(t,t_0)+ \tilde{\mathbf{C}}(t,t_0) \mathbf{W}^\dagger(t) \right]_{i, j} + \Tr\left[ \hat{b}_j^\dagger \hat{b}_i \mathcal{L}'\hat{\rho}(t,t_0)\right] - C_{i, j} \Tr \left[ \mathcal{L}'\hat{\rho}(t,t_0) \right].
\end{align}
The last term may be written compactly in terms of the matrix elements of 
$\mathbf{G}$, such that
\begin{align}
    \Tr\left[ \hat{b}_j^\dagger \hat{b}_i \mathcal{L}'\hat{\rho}(t,t_0) \right] - C_{i, j} \Tr \left[ \mathcal{L}'\hat{\rho}(t,t_0) \right] = - \frac{1}{2} \sum_{k,\ell}G_{k, \ell} 
    \Big[
    \langle \{b_j^\dagger b_i, b_k^\dagger b_\ell\} \rangle - 2\langle b_j^\dagger b_i \rangle \langle b_k^\dagger b_\ell\rangle\Big].
\end{align}
The higher-order expectation values may be expressed in terms of quadratic-terms, given that the density operator is Gaussian and remains Gaussian throughout the dynamics~\cite{Sharma:2015}. Invoking Wick's 
theorem, we have
\begin{align}
    \langle b_j^\dagger b_i b_k^\dagger b_\ell \rangle = \langle b_j^\dagger b_i \rangle\langle b_k^\dagger b_\ell \rangle + \langle b_j^\dagger b_\ell \rangle\langle b_i b_k^\dagger \rangle
\end{align}
for all $i,j,k,\ell$; which leads to
\begin{align}
    \Tr\left[ \hat{b}_j^\dagger \hat{b}_i \mathcal{L}'\hat{\rho}(t,t_0) \right] - C_{i, j} \Tr \left[ \mathcal{L}'\hat{\rho}(t,t_0) \right] &= - \frac{1}{2} \sum_{k,\ell}G_{k, \ell} 
    \Big[ 
    \langle b_j^\dagger b_\ell \rangle\langle b_i b_k^\dagger \rangle + \langle b_k^\dagger b_i \rangle \langle b_\ell b_j^\dagger \rangle
    \Big]
    \\[0.3cm]
    &= - \frac{1}{2} \left[ \mathbf{C}(t) \mathbf{G} (\bm{1}-\mathbf{C}(t)) + (\bm{1}-\mathbf{C}(t))\mathbf{G} \mathbf{C}(t) \right]_{i, j}.
\end{align}
Finally, we therefore arrive at 
\begin{align}
    \frac{\rm{d}\mathbf{\tilde{C}}(t,t_0)}{\rm{d}t} = - \left[\mathbf{W}(t)\mathbf{\tilde{C}}(t,t_0) + \mathbf{\tilde{C}}(t,t_0) \mathbf{W}^\dagger(t) \right] -\frac{1}{2} \Big[ \mathbf{C}(t) \mathbf{G} [\bm{1} - \mathbf{C}(t)] + [\bm{1} - \mathbf{C}(t)] \mathbf{G} \mathbf{C}(t) \Big].
\end{align}
Note that we have suppressed the initial-time dependence in $\mathbf{C}(t)$ but not in $\mathbf{\tilde{C}}(t,t_0)$. This can be done since, from Eq.~\eqref{fcs_average_current}, only the integrated current depends on the initial condition but not the instantaneous current. For the latter, only knowledge of $\mathbf{C}(t)$ is required. For the instantaneous noise, however, it is important to keep track of the initial condition through $\mathbf{\tilde{C}}(t,t_0)$ due to the non-additivity inherent to this quantity. 
In multi-reservoir configurations, the relevant equations maintain the same form but $\mathbf{G}$ changes according to Eq.~\eqref{eq:g_elements}. The only non-zero elements of $\mathbf{G}$ correspond to the $\kappa_k$ system-lead couplings, for the matrix elements that couple the system with the reservoir over which the particle statistics are to be computed.

\section{Noise in two-terminal driven junctions}
\label{sec:noise_w}

\subsection{The $\omega \to 0$ limit}

Following the discussion in the main text, we observe that the integrated noise strongly depends on the driving frequency $\omega$. This can be understood from the covariance term that arises when one is interested in the instantaneous noise integrated over many periods in periodically-driven systems.

Let us consider the example proposed in the main text. A central quantum fermionic system is periodically-driven such that $\hat{H}_{\tt{S}}$ is given by Eq.~10 (main text). The system is driven out of equilibrium by the action of thermal reservoirs kept at different temperatures and chemical potentials. The state of the system $\hat{\rho}_{\tt{S}}(t,t_0)$ becomes periodic after a sufficiently-long time, a condition that we have dubbed \emph{limit cycle} (LC), a notion being reminiscent of a similar notion in the context of synchronisation. Let us define the time in which this condition occurs as $t = t_1$ and at the same time, consider the subsequent periods for counting particles. In the main text, we used the condition $J_{\nu}(t_1 + \tau) = J_{\nu}(t_1)$ to establish this periodicity of the state, where $\tau \defeq 2\pi/ \omega$ and $\nu = \tt{L}, \tt{R}$. Since the current depends only on the state at time $t$ [Eq.~5 (main text)], if $\hat{\rho}_{\tt{S}}(t_1 + \tau,t_1) = \hat{\rho}_{\tt{S}}(t_1, t_1)$, then the associated currents must also be periodic from $t = t_1$ forward. 

We now consider the total charge that flows from one reservoir to the other as a function of time, i.e., $N_{\nu}(t, t_1)$, starting from $t = t_1$ after the LC has been reached. Since the total charge is additive, we may express this quantity as a sum of the accumulated charge over many periods
\begin{align}
    N_{\nu}(t, t_1) &= N_{\nu}(t_1 + \tau, t_1) + N_{\nu}(t_1 + 2\tau, t_1 + \tau) + \cdots + N_{\nu}(t_1 + a\tau, t_1 + [(a-1)\tau]) \nonumber \\
    &= \sum_{a=1}^{m} N_{\nu}(t_1 + a\tau, t_1 + [(a-1)\tau]),
\end{align}
where we have parametrised $t = t_1 + m\tau$. The crucial aspect to be highlighted is that to understand fluctuations, we must consider the variance of $N_{\nu}(t, t_1)$. We have
\begin{align}
    \vvar{[N_{\nu}(t, t_1)]} = \vvar{\left[ \sum_{a=1}^{m} N_{\nu}(t_1 + a\tau, t_1 + [(a-1)\tau]) \right]}.
\end{align}
Defining $N^a_{\nu} \defeq N_{\nu}(t_1 + a\tau, t_1 + [(a-1)\tau])$, we may express the variance as
\begin{align}
    \vvar{[N_{\nu}(t, t_1)]} &= \sum_{a,b=1}^{m} \ccov{[N^a_{\nu},N^b_{\nu}]} \nonumber \\
    &= \sum_{a=1}^{m} \vvar{[N^a_{\nu}]} + \sum_{a \neq b} \ccov{[N^a_{\nu},N^b_{\nu}]}.
\end{align}
In the LC, the accumulated charge per cycle is equivalent for each cycle after $t=t_1$, then 
\begin{align}
\label{eq:var_cov_sum}
    \vvar{[N_{\nu}(t, t_1)]} = \vvar{[N_{\nu}(t_1 + m\tau, t_1)]} = m\vvar{[N_{\nu}(t_1 + \tau, t_1)]} + \sum_{a \neq b} \ccov{[N^a_{\nu},N^b_{\nu}]}.
\end{align}
It is important to stress that the above result does not depend on $t_1$, so long as it remains a timescale large enough such that the LC condition is reached. 

The correspondence between $\overline{S^{0}_{\nu}}$ and $\overline{S^{\infty}_{\nu}}$ in the $\omega \to 0$ regime becomes further manifest by first noting that the \emph{zero-frequency component} of the noise, denoted by $\overline{S^{\infty}_{\nu}}$, can be defined in \emph{two equivalent forms}. The first corresponds to definition in Eq.~13 (main text), i.e., as the the integral over a single period in the limit of long time of the noise after the limit cycle
\begin{align}
\label{eq:s_infty_bar}
    \overline{S^{\infty}_{\nu}} \defeq \lim_{t\to\infty}\frac{1}{\tau} \int_{0}^{\tau} {\rm d}t' D_\nu(t+t',t_1),
\end{align}
while the second equivalent form is via the total time-average of the noise over the entire time domain after the limit cycle
\begin{align}
\label{eq:s_infty_bar_standard}
    \overline{S^{\infty}_{\nu}} \defeq \lim_{t\to\infty}\frac{1}{t} \int_{t_1}^{t} {\rm d}t' D_\nu(t',t_1),
\end{align}
with $D_{\nu}(\bullet,\bullet)$ being defined in Eq.~6 (main text). Both these definitions are equivalent in systems with a periodic time-dependent Hamiltonian given that the noise itself becomes periodic in the limit of long time. Periodicity guarantees that these two definitions are equivalent in the $t \to \infty$ limit. To visualise this, one may go back to Fig.~2 (bottom)(main text), where it becomes apparent that in the limit of long time whereby the instantaneous noise is periodic, the average value over a single period [dark grey region on the right side of Fig.~2 (bottom)(main text)] is approximately the same as the time average of the entire time interval starting from $t = t_1$. Such approximation becomes arbitrarily accurate as more time periods are considered. 

In the main text, we considered the first definition highlighted in Eq.~\eqref{eq:s_infty_bar} as it seems more natural in our context, although, the zero-frequency component of the noise as defined in Eq.~\eqref{eq:s_infty_bar_standard} is the one most-commonly found in the literature~\cite{Hanggi2003}. We introduce here the total-time average definition [Eq.~\eqref{eq:s_infty_bar_standard}] since it helps us understand the $\omega \to 0$ limit from our statistical reasoning about the variances described above.

We now may understand the $\omega \to 0$ limit, in which $\tau \to \infty$. In Eq.~\eqref{eq:var_cov_sum}, if we set $m = 1$, the covariance term should not be included and, taking the $t \to \infty$ time-average on the left-hand side yields
\begin{align}
    \overline{S^{\infty}_{\nu}} &= \lim_{t \to \infty} \frac{\vvar{[N_{\nu}(t, t_1)]}}{t} \nonumber \\
    &= \lim_{\tau \to \infty}\frac{\vvar{[N_{\nu}(t_1 + \tau, t_1)]}}{\tau} \nonumber \\
    &= \lim_{\tau \to \infty}\frac{1}{\tau} \int_{0}^{\tau} {\rm{d}}t'D_{\nu}(t_1 + t', t_1) = \lim_{\tau \to \infty} \overline{S^{0}_{\nu}}.
\end{align}
Consequently, for systems with autonomous non-equilibrium steady states or periodically-driven systems with very slow driving frequencies, it is of no consequence to associate a difference between the definitions of the charge fluctuations. However, in systems with sufficiently fast driving frequencies, there exists a fundamental difference as highlighted in Eq.~\eqref{eq:var_cov_sum} and Fig.~3 (main text).

\subsection{Asymmetric coupling}

Our discussion thus far has been focused on the particular case of a symmetric configuration. An in the main text, the effective system-reservoir coupling $\Gamma_{\tt L} = \Gamma_{\tt R} = \Gamma$ is identical for both reservoirs, the mean chemical potential was set to 
\begin{equation}
\overline{\mu} \defeq (\mu_{\tt L} + \mu_{\tt R}) / 2 = 0 
\end{equation}
and the temperatures were set equal $T_{\tt L} = T_{\tt R} = T$. The system Hamiltonian $\hat{H}_{\tt S}$ in Eq.~10 (main text) is also symmetric with respect to the reservoirs. If one relaxes any of these conditions, the instantaneous currents will not, in general, coincide at any point in time, i.e., $J_{\tt L}(t) \neq J_{\tt R}(t)$, and neither will the noise after the limit cycle $D_{\tt L}(t,t_1) \neq D_{\tt R}(t,t_1)$ where $t = t_1$ an arbitrary point in time in which the instantaneous current (either from the left or the right reservoir) becomes periodic. 

\begin{figure}[b]
\fontsize{13}{10}\selectfont 
\centering
\includegraphics[width=0.49\columnwidth]{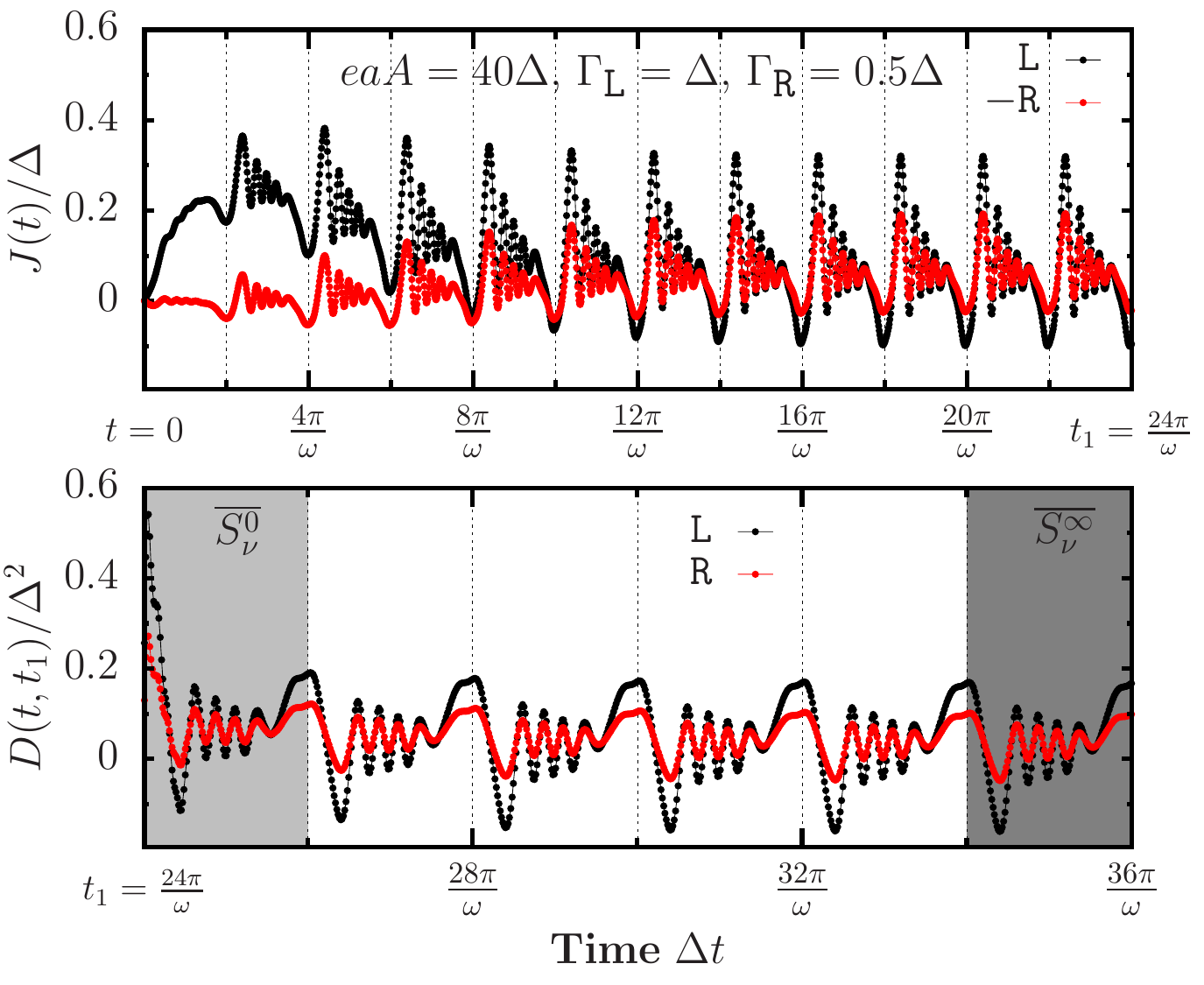} \includegraphics[width=0.49\columnwidth]{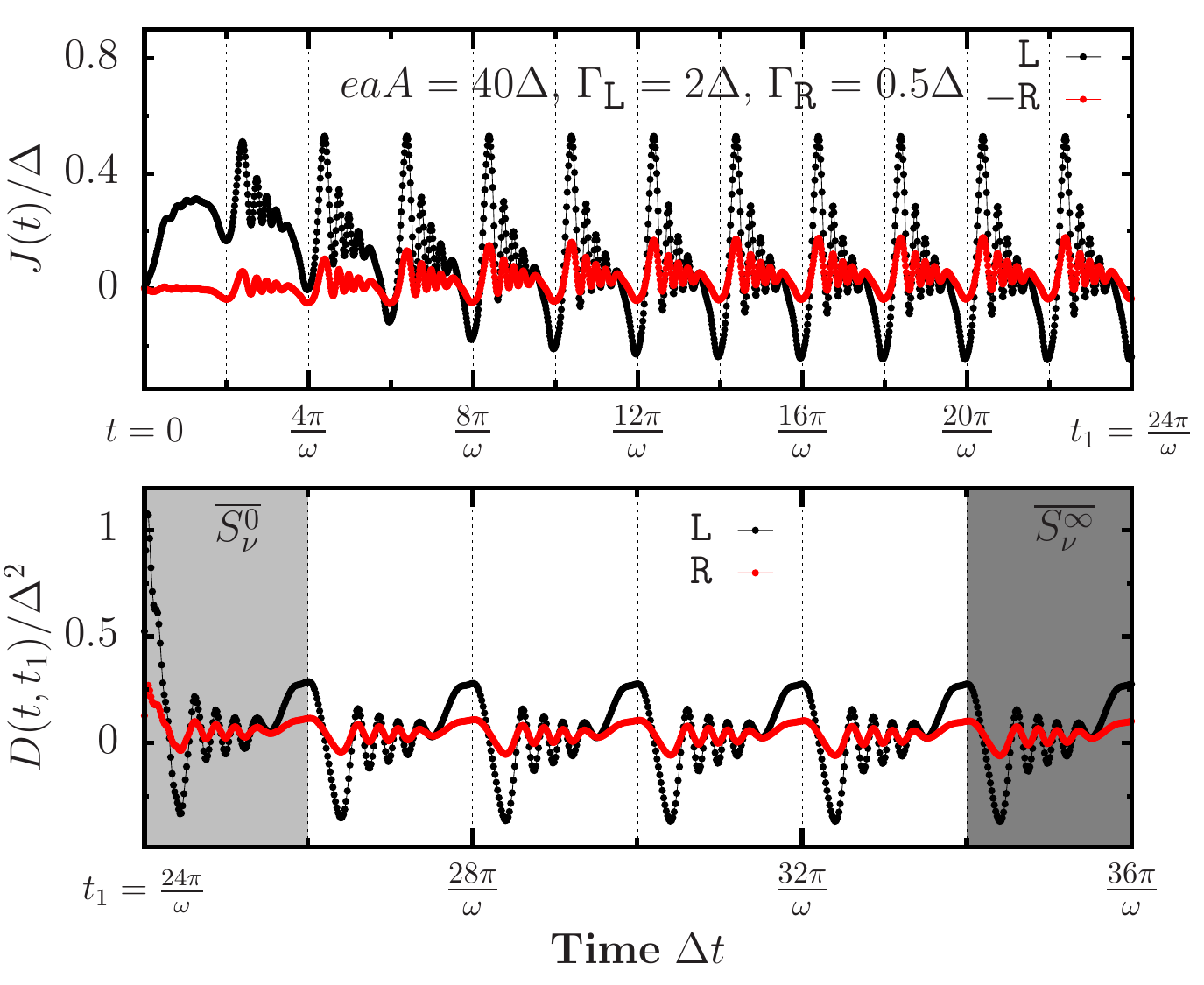}
\caption{(Top) Average currents $J_{\tt L}(t)$ and $-J_{\tt R}(t)$ [Eq.~5 (main text)] for multiple periods $\tau=2\pi/\omega$ of the external drive, up until the LC is reached for asymmetric system-reservoir coupling strengths  (left) $\Gamma_{\tt L} = \Delta$ and (right) $\Gamma_{\tt L} = 2\Delta$. (Bottom) Corresponding noise $D_{\tt L/R}(t,t_1)$ for both asymmetric system-reservoir coupling cases, starting at the LC $t_1 = 24\pi/\omega$. Integrating over first period yields $\overline{S^{0}_{\nu}}$ in Eq.~12 (main text).
Waiting for multiple periods and then integrating yields instead $\overline{S^{\infty}_{\nu}}$ in Eq.~13 (main text). In these simulations we set $\Gamma_{\tt R} = 0.5\Delta$, $\omega = 5\Delta$, $T_{\tt L} = T_{\tt R} = T = 0.1\Delta$, $\mu_{\tt L} = -\mu_{\tt R} = 24\Delta$, $eaA = 40\Delta$. Reservoir parameters are the same as in the main text, $\gamma_{k, \alpha} = 2W/N$, $\kappa_{k,p} = \sqrt{\Gamma_{\alpha}\gamma_{k, \alpha}/2\pi}$ and $W = 100\Delta$ with $N = 400$, with $\Delta$ the hopping parameter in the system Hamiltonian $\hat{H}_{\tt S}$ in Eq.~10 (main text).}
\label{fig:4}
\end{figure}

This can be observed explicitly in Fig.~\ref{fig:4}, where we have relaxed the condition of symmetric coupling to the reservoirs by computing the current and noise with both $\Gamma_{\tt L} = \Delta$ [Fig.~\ref{fig:4}  (left)] and $\Gamma_{\tt L} = 2\Delta$ [Fig.~\ref{fig:4} (right)] at fixed $\Gamma_{\tt R} = 0.5\Delta$. The average currents $J_{\tt L/R}(t)$ do not yield the same amplitude under these conditions, yet, the LC condition still holds for sufficiently long time such that $J_{\nu}(t + \tau) = J_{\nu}(t)$. Using our previous convention we shall define $t = t_1$ the point in time in which this condition is satisfied. Note that this condition appears to hold even before the selected value of $t_1 = 24\pi/\omega$. This choice is arbitrary as long as $J_{\nu}(t + \tau) \approx J_{\nu}(t)$. 

Interestingly, however, it appears that in both cases shown in Fig.~\ref{fig:4} (top), the LC-averaged current  
\begin{equation}
    \overline{J}_\nu = \frac{1}{\tau} \int_{t_1}^{t_1+\tau} {\rm d} t^{\prime} J_\nu(t^{\prime})
\end{equation}
computed from either the left or right reservoirs yields the same average (with a sign difference). We have confirmed numerically that this is indeed the case. Such condition is nothing but conservation of charge over one period in the LC, in which the accumulated charge over one period is given by the expectation value of $N_{\nu}(t_1 + \tau,t_1)$, given by $\langle N_{\nu}(t_1 + \tau,t_1) \rangle = \int_{t_1}^{t_1 + \tau} {\rm{d}}t' J_{\nu}(t')$.

With our sign convention, where the current is positive if flowing from $\tt L$ to $\tt R$, the LC-averaged currents cancel each other $\overline{J}_{\tt L} + \overline{J}_{\tt R} = 0$. It immediately follows that the accumulated charge over one period in the LC is equivalent, up to a minus sign, computed from either $\tt L$ or $\tt R$, i.e., $\overline{N} = \langle N_{\tt L}(t_1 + \tau,t_1) \rangle + \langle N_{\tt R}(t_1 + \tau,t_1) \rangle = 0$. 

The charge $N_{\nu}(t,t_1)$ is a random variable. However, it is clear from the above argument that $N_{\tt L}(t,t_1)$ and $N_{\tt R}(t,t_1)$ cannot be independent random variables and must somehow be correlated, at least in general. The first few periods in Fig.~\ref{fig:4} (bottom) indicate that this is indeed the case from the dynamics of $D_{\nu}(t,t_1)$. 

This can be understood from the perspective of the configuration as a whole, where system and baths form the universe evolving in time under unitary dynamics. From this perspective, the total charge divided among each sub-partition must satisfy
\begin{align}
    \langle N_{\tt S}(t, t_1) \rangle + \langle N_{\tt L}(t, t_1) \rangle + \langle N_{\tt R}(t, t_1) \rangle = 0,
\end{align}
at any point in time $t$ since the total charge is conserved in the global configuration. It follows that the variance of the accumulated charge computed from either the perspective of the $\tt L$ or $\tt R$ reservoir is 
correlated to the other reservoir through the system
\begin{align}
\label{eq:var_sys_left_right}
    \vvar{[N_{\tt L}(t, t_1)]} &= \vvar{[-N_{\tt S}(t, t_1) - N_{\tt R}(t, t_1)]} \nonumber \\
    &= \vvar{[N_{\tt S}(t, t_1)]} + \vvar{[N_{\tt R}(t, t_1)]} + 2\ccov{[N_{\tt S}(t, t_1), N_{\tt R}(t, t_1)]},
\end{align}
and, naturally, so is the noise
\begin{align}
    D_{\tt L}(t,t_1) = D_{\tt S}(t,t_1) + D_{\tt R}(t,t_1) + 2\frac{{\rm d}}{{\rm d}t} \ccov{[N_{\tt S}(t, t_1), N_{\tt R}(t, t_1)]}.
\end{align}
A very special case consists of considering a symmetric configuration, such as we did in the main text. In this case, after $t = t_1$ where the LC cycle has been reached, the instantaneous current from $\tt L$ or $\tt R$ are equivalent, at any point in time. The condition $J_{\tt L}(t) = J_{\tt R}(t)$ [Fig.~2 (bottom)(main text)] implies that $N_{\tt L}(t,t_1) = N_{\tt R}(t,t_1)$ and $N_{\tt S}(t,t_1) = 0$, i.e., the accumulated charge through the system is always zero $\forall t>t_1$. This is not the case in general, where we have instead that this condition holds only on average over a single period after $t = t_1$. In this particular case, from Eq.~\eqref{eq:var_sys_left_right}
\begin{align}
    \vvar{[N_{\tt L}(t, t_1)]} = \vvar{[N_{\tt R}(t, t_1)]} \nonumber \\
    \implies D_{\tt L}(t,t_1) = D_{\tt R}(t,t_1).
\end{align}
This is, however, only a special case and further remarks the subtleties associated to the operational definitions of the noise as we highlight in our work.

\subsection{Different regimes of operation}

\begin{figure}[t]
\fontsize{13}{10}\selectfont 
\centering
\includegraphics[width=0.49\columnwidth]{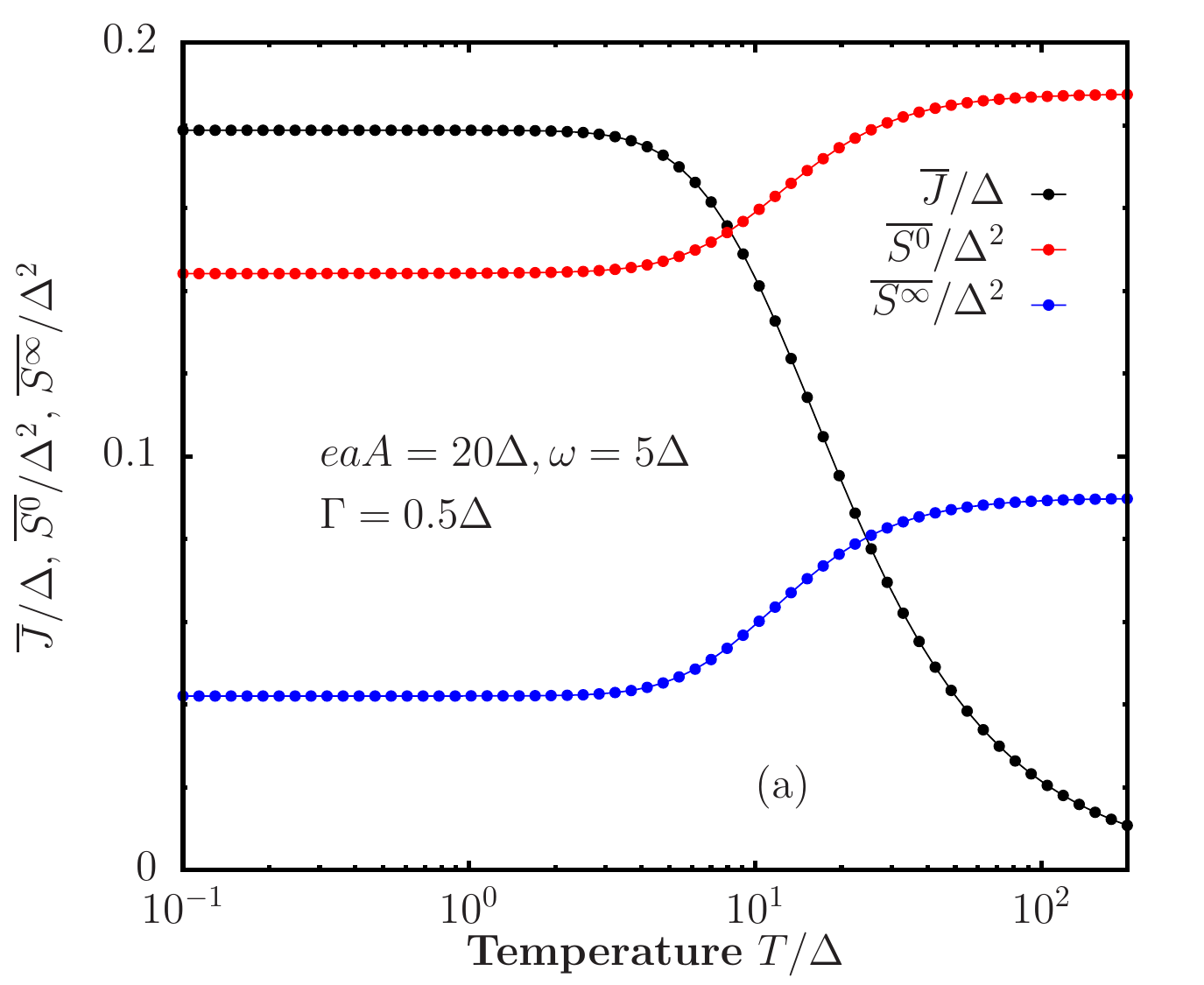} \includegraphics[width=0.49\columnwidth]{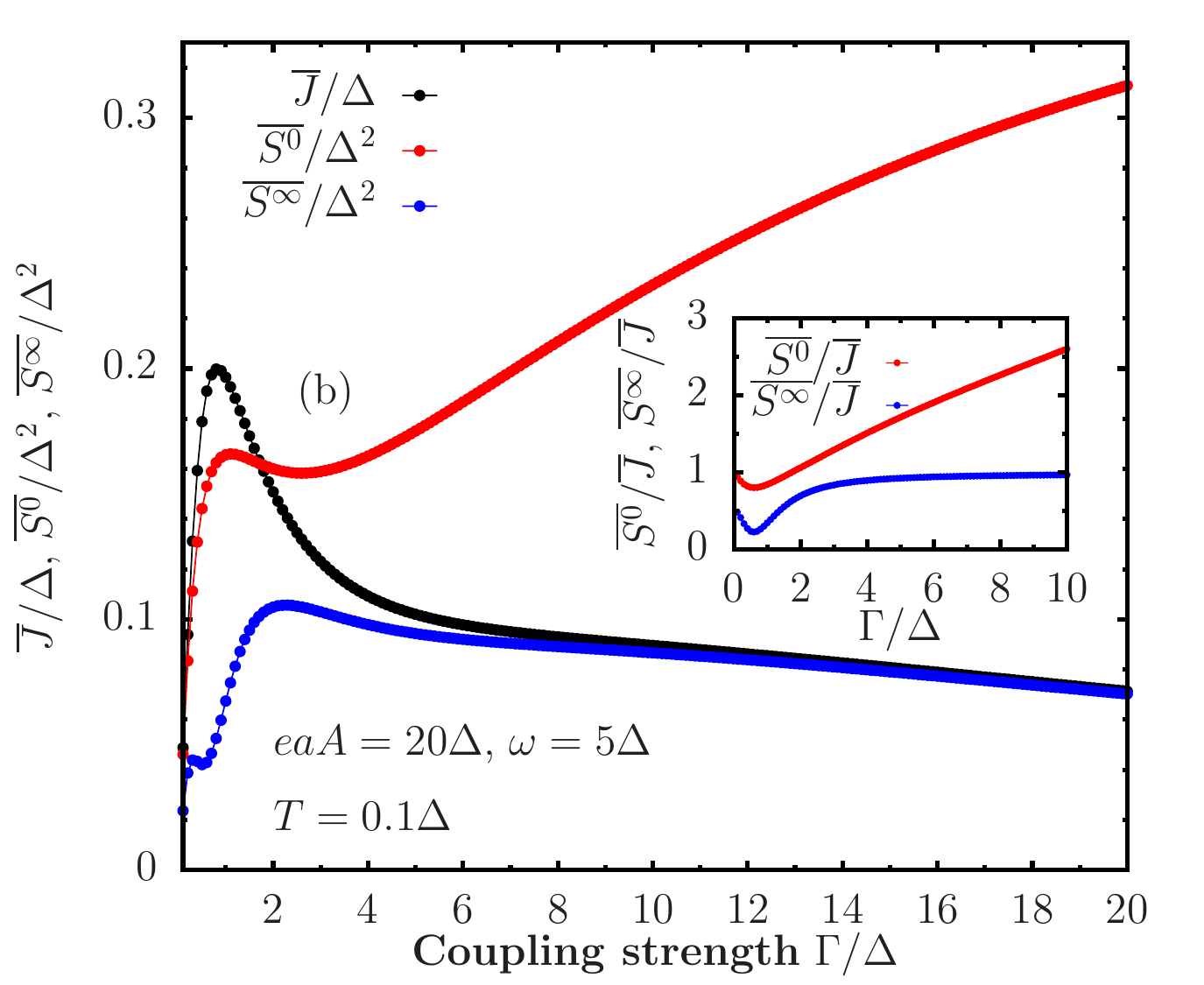}
\caption{$\overline{J}$, $\overline{S^{0}}$ and $\overline{S^{\infty}}$ [Eqs.~11-13 (main text)] as a function of the (a) temperature $T = T_{\tt L} = T_{\tt R}$ and (b) the system-reservoir coupling $\Gamma = \Gamma_{\tt L} = \Gamma_{\tt R}$. Calculations shown for a fixed driving field $eaA = 20\Delta$, driving frequency $\omega = 5\Delta$ and chemical potential $\mu_{\tt L} = -\mu_{\tt R} = 24\Delta$. Reservoir parameters are the same as in the main text, $\gamma_{k, \alpha} = 2W/N$, $\kappa_{k,p} = \sqrt{\Gamma_{\alpha}\gamma_{k, \alpha}/2\pi}$ and $W = 100\Delta$ with $N = 400$, with $\Delta$ the hopping parameter in the system Hamiltonian $\hat{H}_{\tt S}$ in Eq.~10 (main text). The inset in (b) displays the ratio of the variance at the limit cycle $\overline{S^0}$ and the zero-frequency component of the noise $\overline{S^{\infty}}$, and the LC-averaged current $\overline{J}$ for the same parameters as in the main panel.}
\label{fig:5}
\end{figure}

The goal of this section is to expose the dynamics of the current and its noise in the time-dependent model $\hat{H}_{\tt S}$ written in Eq.~10 (main text) within different regimes of temperature $T$ and system-reservoir coupling $\Gamma$ than those shown in the main text. Fig.~\ref{fig:5}(a) shows the averaged current over a single period in the LC ($\overline{J}$) the variance over one period in the LC ($\overline{S^{0}}$) and the zero-frequency component of the noise ($\overline{S^{\infty}}$) as a function of the temperature $T = T_{\tt L} = T_{\tt R}$. For these calculations we fixed $\Delta$, the frequency of the driving $\omega = 5\Delta$, the driving field strength $eaA = 20\Delta$, the chemical potential $\mu_{\tt L} = -\mu_{\tt R} = 24\Delta$ and the effective system-reservoir coupling $\Gamma = \Gamma_{\tt L} = \Gamma_{\tt R} = 0.5\Delta$. As stated in the main text, the reservoirs are parametrised using a finite number of modes $N = 400$ which guaranteed convergence of all three quantities. Fig.~\ref{fig:5}(a) reveals that the calculations presented thus far represent the zero-temperature regime of operation. Such is the case as it can be observed that for $T \lesssim \Delta$, neither $\overline{J}$, $\overline{S^{0}}$ or $\overline{S^{\infty}}$ change below this threshold. An important remark to be made is that the zero-temperature regime is the most difficult to address with the mesoscopic leads approach. This can be understood from the behaviour of the Fermi-Dirac distribution in the zero-temperature limit which becomes discontinuous at $\mu_{\alpha}$. This implies that the transport will be probed over very small energy scales and the discretisation of the reservoir would be more prominent in this regime. However, even so we can reliably compute the transport and its noise with fidelity with a tractable number of modes. At higher temperatures, fewer modes $N$ are required to attain convergence. 

Varying the coupling to the reservoirs $\Gamma = \Gamma_{\tt L} = \Gamma_{\tt R}$ results in non-monotonic behaviour of the average current and its concomitant average noise at fixed temperature $T = 0.1\Delta$. This effect is highlighted in Fig.~\ref{fig:5}(b) where there exists a value of $\Gamma$ for which the average current reaches a maximum value. With respect to the average noise, we observe in Fig.~\ref{fig:5} that there is a stark contrast of the fluctuations depending on how they are operationally defined over time periods. The inset in Fig.~\ref{fig:5}(b) displays the ratio $\overline{S^{\infty}} / \overline{J}$, also defined as the Fano factor in the literature~\cite{Blanter2001,Hanggi2003,Hanggi2004}. In the regime of high voltage, we have that if the coupling to the reservoirs $\Gamma \ll \Delta$, then each of the two contacts act like a transport barrier. This regime can be understood as a static double-barrier configuration and in this limit the shot noise $\overline{S^{\infty}} / \overline{J} \approx 1/2$. Alternatively, in the opposite regime whereby $\Gamma \gg \Delta$, it is the link between the two fermionic sites that acts a single barrier, in which $\overline{S^{\infty}} / \overline{J} \approx 1$~\cite{Blanter2001,Hanggi2004}. These limits only hold at zero-temperature. Our approach provides no limitations in the regimes of operation described by the voltage, coupling to reservoirs, driving frequency or temperature regimes. It is reassuring, however, to recover known limits for the model at hand. We observed from the inset in Fig.~\ref{fig:5}(b) that the variance over one period in the LC, $\overline{S^{0}}$, behaves very differently. In particular, in the limit $\Gamma \gg \Delta$, the variance to average current ratio keeps increasing monotonically as a function of $\Gamma$. The difference stems from the very definition of the fluctuations as we highlight in the main text, where we find that both of these definitions yield the same results in the $\omega \to 0$ limit.

\end{document}